\documentclass[10pt,aps,prl,twocolumn,superscriptaddress]{revtex4-2}

\usepackage[pages=all, color=black, position={current page.south}, placement=bottom, scale=1, opacity=1, vshift=5mm]{background}
\SetBgContents{
	\tt 
}      

\usepackage[margin=1in]{geometry} 

\usepackage{amsmath}
\usepackage{amsthm}
\usepackage{amssymb}
\usepackage{times}
\usepackage{braket}
\usepackage{ulem}

\usepackage{multirow}
\usepackage{array}
\usepackage{makecell}
\usepackage{diagbox}
 \usepackage{booktabs}

\usepackage{tikz}
\usetikzlibrary{quantikz2}

\usepackage[utf8]{inputenc}
\usepackage{svg}
\usepackage{hyperref}

\usepackage[english]{babel}

\usepackage{graphicx, color}
\graphicspath{{fig/}}

\usepackage{algorithm, algpseudocode} 
\usepackage{mathrsfs} 
\usepackage{lipsum}
\newcommand{\drangle}{\rangle\!\rangle}

\usepackage[dvipsnames]{xcolor}

\begin{document}

\title{High-efficiency photonic GHZ measurements with linear optics}

\author{Stefan N. van den Hoven}
\affiliation{MESA+ Institute for Nanotechnology, University of Twente, 7500AE Enschede, The Netherlands}

\author{Love A. Pettersson}
\thanks{Present address, Sygaldry Technologies Denmark ApS}
\affiliation{Center for Hybrid Quantum Networks (Hy-Q), Niels Bohr Institute, University of Copenhagen, Jagtvej 155A, 
2200 Copenhagen N, Denmark}
\affiliation{NNF Quantum Computing Programme, Niels Bohr Institute, University of Copenhagen, Blegdamsvej 17, 2100 Copenhagen, Denmark.}

\author{Anders S. Sørensen}
\affiliation{Center for Hybrid Quantum Networks (Hy-Q), Niels Bohr Institute, University of Copenhagen, Jagtvej 155A, 
2200 Copenhagen N, Denmark}

\author{Stefano Paesani}
\affiliation{NNF Quantum Computing Programme, Niels Bohr Institute, University of Copenhagen, Blegdamsvej 17, 2100 Copenhagen, Denmark.}

\date{
	\today
}

\begin{abstract}
Probabilistic and destructive Bell-measurements, also known as fusions, constitute a fundamental primitive for photonic quantum technologies. These operations rely only on pairwise photonic interference, whereas the computational power of photonic quantum systems originates from many-body interference. Here, we introduce a family of high-efficiency linear-optical GHZ measurements based on four-mode interferometers and show that incorporating many-body interference into this ubiquitous primitive leads to more efficient entangling operations. For the four-photon case, the success probability is $9/32$, and adaptive variants ensure that unsuccessful events project onto useful stabilizer states. We further show that the approach naturally generalizes to larger GHZ measurements, yielding an exponentially higher success probability than previous linear-optical schemes. Finally, we demonstrate practical advantages for graph-state resource generation, entanglement distribution between remote spin-photon interfaces, and fusion-based quantum computing. Our results establish high-efficiency many-body measurements as a new primitive for scalable photonic quantum information processing.
\end{abstract}

\maketitle


\section{Introduction}
Joint measurements of photons represent a central tool for generating and processing entanglement in quantum photonic systems. 
The non-linear nature of quantum measurements has emerged as a viable approach to make entangling operations between photons in linear optics~\cite{knill2001scheme, PhysRevLett.95.010501, PhysRevA.68.032310}, i.e., through evolutions based on interferometers composed of beam-splitters and phase-shifters. 
Because of the lack of direct photon-photon interactions, entanglement generation between individual photons is otherwise impossible with linear optics.
Schemes for linear-optical Bell measurements, i.e., a projection into a two-photon entangled basis state, have therefore been studied extensively.
They play a central role in many quantum technologies, including quantum repeaters~\cite{PhysRevLett.81.5932, RevModPhys.95.045006}, quantum networks~\cite{Kimble2008, doi:10.1126/science.aam9288}, and quantum computer architectures~\cite{bartolucci2023fusion,chan2025practicalblueprintlowdepthphotonic,PRXQuantum.6.020304, wein2024minimizingresourceoverheadfusionbased, ostmann2025nonlinearphotonicarchitecturefaulttolerant}. In the latter, these Bell measurements are typically referred to as fusion operations because they are used to fuse entangled states together into larger structures.

A well-known challenge for linear-optical Bell measurements of two photons is that they are intrinsically probabilistic, with a limit to the success probability of 50\%~\cite{Calsamiglia2001}.
This probability can be boosted arbitrarily close to unity through the use of additional resources, for example up to 75\% by adding 4 individual photons~\cite{PhysRevLett.113.140403} or to $1- 1/2^{n+1}$ by adding increasingly large entangled ancillae~\cite{PhysRevA.84.042331}, consisting of up to $2^n$ photons.
Bell measurements, however, fundamentally rely on the interference of only two photons. By contrast, the computational power of photonic quantum systems is widely understood to originate from genuine multiphoton interference~\cite{Aaronson_Arkhipov_BS,renema2018efficient, Oh:25} which has recently been used to improve tasks such as photonic state characterization and photon distillation~\cite{hoven2026quantumadvantagesinglephotonstate, schadow2026certificationlinearopticalquantum,somhorst2024photon}. This naturally raises the question whether the ubiquitous Bell measurement can be generalized to exploit multiphoton interference.
We answer this question affirmatively by introducing a family of high-efficiency linear-optical GHZ measurements based on four-mode interferometers. For four photons, our scheme increases the success probability by a factor of $9/4$ compared to the schemes presented in Refs.~\cite{PhysRevA.110.032402, Saied2024AQN, bartolucci2021creation}. We further show that the approach generalizes to larger GHZ measurements, yielding an exponentially higher success probability, and demonstrate advantages for graph-state generation, entanglement distribution, and fusion-based quantum computing.


\section{High-efficiency four-body GHZ measurement} 
We begin with the smallest case for which our multiphoton scheme applies an improvement over previous schemes: a four-body GHZ measurement, meaning a successful projection onto a state that has the following stabilizer generators: $\{\pm X_1X_2X_3X_4,\pm Z_1Z_2, \pm Z_2Z_3, \pm Z_3Z_4\}$ e.g. $1/\sqrt{2}(|0000\rangle+|1111\rangle$). Here, $X_i$ and $Z_i$ denote the Pauli operators acting on qubit $i$, and a stabilizer is an operator for which the state is an eigenstate with a definite eigenvalue~\cite{gottesman1997stabilizer, gottesman1998heisenbergrepresentationquantumcomputers}. We consider destructive and probabilistic 4-GHZ measurements, which generalize destructive and probabilistic Bell measurements (type-II fusions~\cite{PhysRevLett.95.010501}).

The success probability of a scheme corresponds to the probability that a detection event can be coupled to a unique 4-GHZ state. For calculating the success probability we consider a situation where the incident photons have equal probabilities of occupying all possible input states when tracing over photons to which they are entangled, i.e. the reduced density matrix of the incident photons is the identity.

Previous linear-optical schemes for four-body GHZ measurements succeed only when two of the sixteen GHZ states can be uniquely identified, limiting the success probability to $1/8$~\cite{PhysRevA.110.032402, Saied2024AQN,bartolucci2021creation}. By exploiting many-body interference, our design increases this success probability by a factor of $9/4$.

A successful n-GHZ projection consists of determining all of its stabilizer generators. The sign in front of the XXXX generator translates to the sign between the two terms of a 4-GHZ state. Thus, determining the sign of the XXXX generator requires that both parts of the superposition can result in the same heralding pattern. With a correctly chosen interferometer, these paths interfere either constructively or completely destructively depending on the sign between the terms of the superposition. Thus, detection events are linked to the sign of the XXXX generator.

\begin{figure*}
    \centering
    \includegraphics[width=1.0\linewidth]{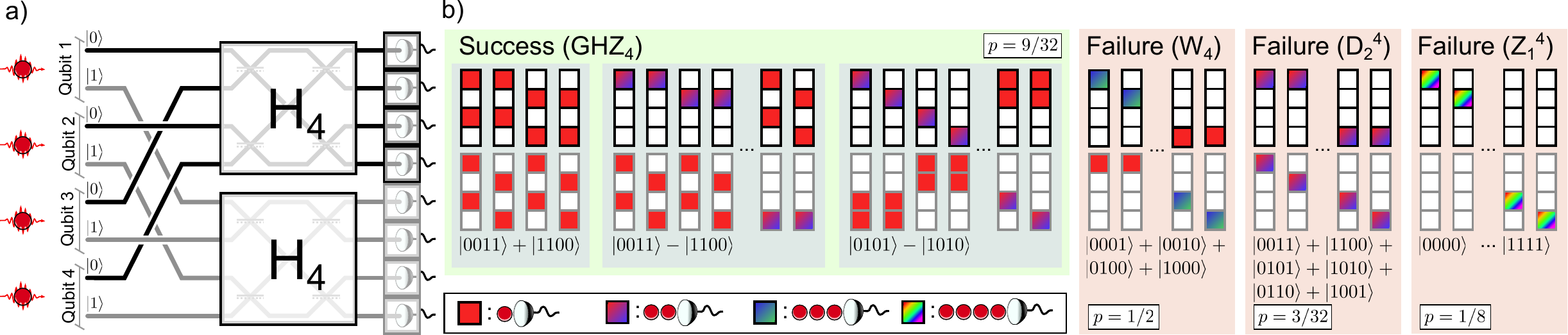}
    \caption{High-efficiency four-body GHZ measurement. a) Linear-optical network that divides dual-rail photonic qubit over two $H_4$ interferometers. Upon detecting two photons in each subsystem such that at least one subsystems anti-bunches, a projection onto a $4$-GHZ is performed. b) Details of heralding patterns and the corresponding projection in qubit space. }
    \label{fig1: GHZ projection}
\end{figure*}

This motivates an interferometer consisting of two subsystems, with one rail from each photonic dual-rail-encoded qubit directed to each subsystem, as depicted in Fig.~(\ref{fig1: GHZ projection}a). Each GHZ-state that divides the photons equally over both subsystems does so for both terms in its superposition, enabling interference that could yield information about the XXXX stabilizer. Six of the 16 possible 4-GHZ states divide the photons equally over both subsystems. 

Each subsystem implements an interferometer whose transition amplitudes have equal magnitudes and phases of 0 or $\pi$. This ensures constructive or completely destructive interference between the superposition terms of the GHZ state. This is realized by the four-mode Hadamard:

\begin{equation}
H_4=\frac{1}{2}
    \begin{pmatrix}
    1&1&1&1\\
    1&-1&1&-1\\
    1&1&-1&-1\\
    1&-1&-1&1\\
\end{pmatrix},
\end{equation}
where the $ij$-th element gives the transition amplitude for a single photon from input mode $i$ to output mode $j$.

Next to the sign of the XXXX stabilizer, we want to determine the signs of all pairwise Z stabilizers. Back-propagating the two-photon Fock-projectors through the interferometer reveals that the four-mode Hadamard matrix also accomplishes this purpose. The detection of two anti-bunched photons (i.e. with mode occupation lists: 1100, 1010, 1001 etc.) corresponds to a projection onto a subspace spanned by the eigenvectors of the three independent pairwise Z stabilizer generators associated with the qubits whose rails are linked to this subsystem. Thus, detecting two photons in an anti-bunching configuration, enables a deduction of the signs of the pairwise stabilizers of the qubits that are featured in this subsystem. However, the detection of two bunched photons (2000, 0200 etc.) corresponds to a projection onto a subspace that is spanned by all states that result in two photons in the respective subsystem and thus does not provide information about the signs of the pairwise stabilizers of the qubits that are featured in this subsystem.

For a successful projection onto a 4-GHZ state, two requirements must be met. First, the detection of two photons in each subsystem. This happens for six out of the total 16 4-GHZ states. Second, we require at least one subsystem to detect the two photons in an anti-bunched configuration. For each GHZ-state that results in two photons in a subsystem, the probability of detecting a bunched configuration is equal to the probability of detecting an anti-bunched configuration. Considering two subsystems, only if both subsystems bunch, we do not retrieve the pairwise Z stabilizers. Hence, the probability of success is given by $\frac{6}{16} \times \frac{3}{4} =\frac{9}{32}$.

Unsuccessful events nevertheless often yield genuinely entangled states. A projection onto a state that is locally Pauli-equivalent to the 4-qubit $W$-state ($W_4=1/2\left[|0001\rangle+|0010\rangle +|0100\rangle +|1000\rangle\right]$) is performed whenever one and three photons are detected in the two subsystems respectively, which occurs with probability $\frac{1}{2}$. A projection onto a state that is locally Pauli-equivalent to the $(4,2)-$Dicke state ($D^{(2)}_4=1/\sqrt{6}[|0011\rangle+|1100\rangle +|0101\rangle +|1010\rangle+|0110\rangle+|1001\rangle]$) is performed whenever two bunching photons are detected in both subsystem, which happens with probability $\frac{3}{32}$. Lastly, the setup projects all four qubits in the single qubit Z basis whenever four photons are detected in either of the subsystems, which happens with probability $\frac{1}{8}$. Figure (\ref{fig1: GHZ projection}b) illustrates examples of qubit projections matched to multiphoton detection events.

Projections onto the $W_4$ and the $D^{(2)}_4$ states could have interesting applications, for example the generation of the $W_4$ state out of four Bell pairs. However, for the applications we have in mind, namely resource state generation or the design of fusion networks for photonic quantum computing~\cite{bartolucci2025comparisonschemeshighlyloss,wein2024minimizingresourceoverheadfusionbased,bartolucci2023fusion,chen2025fusionbasedimplementationqldpccodes}, it is likely more favorable to work with projections onto stabilizer states. 

We therefore present an adaptive strategy to ensure a projection onto a stabilizer state. We detect the first subsystem before the second. If the first subsystem detects two photons in an anti-bunched configuration, the second subsystem is set to an $H_4$ interferometer. For any other detection pattern, the second subsystem switches to two parallel beam splitters. In this way, we project onto a 4-GHZ state with probability $\frac{3}{16}$, we project onto two Bell states with probability $\frac{1}{8}$, we project onto one Bell state with probability $\frac{1}{2}$ and we project in the computational basis with probability $\frac{3}{16}$. In the Supplementary Material~\cite{SM1}, we illustrate this adaptive scheme as well as examples of matches between qubit projections and multiphoton detection events.

Alternatively, we can switch the second subsystem to the identity matrix  $I_4$. We then project onto a 4-GHZ state with probability $\frac{3}{16}$ and in the computational basis with probability $\frac{13}{16}$. Alternative adaptive strategies are possible, which can balance the probability of projecting onto a 4-GHZ state and the probability of projecting on a stabilizer state in case of failure; we have highlighted extreme cases here. 

\section{Schemes for multi-body GHZ measurement}
Existing linear optical schemes that allow for multi-body GHZ projections~\cite{PhysRevA.110.032402, Saied2024AQN, bartolucci2021creation} have a low success probability because only $2$ of $2^n$ GHZ states can be projected onto, as in the 4-GHZ case. Naturally, we ask whether we can generalize our scheme for 4-GHZ projections to the $n$-qubit case. Two generalizations arise. The first splits up the components of dual rail qubits into two subsystems which feature large Hadamard matrices. The second introduces more $H_4$ subsystems and distributes the dual-rail qubits over these subsystems. The former generalization makes it challenging to retrieve the pairwise Z stabilizers. We therefore leave further investigation of the former generalization to future work and proceed with the latter.

\begin{figure}
    \centering
    \includegraphics[width=1\linewidth]{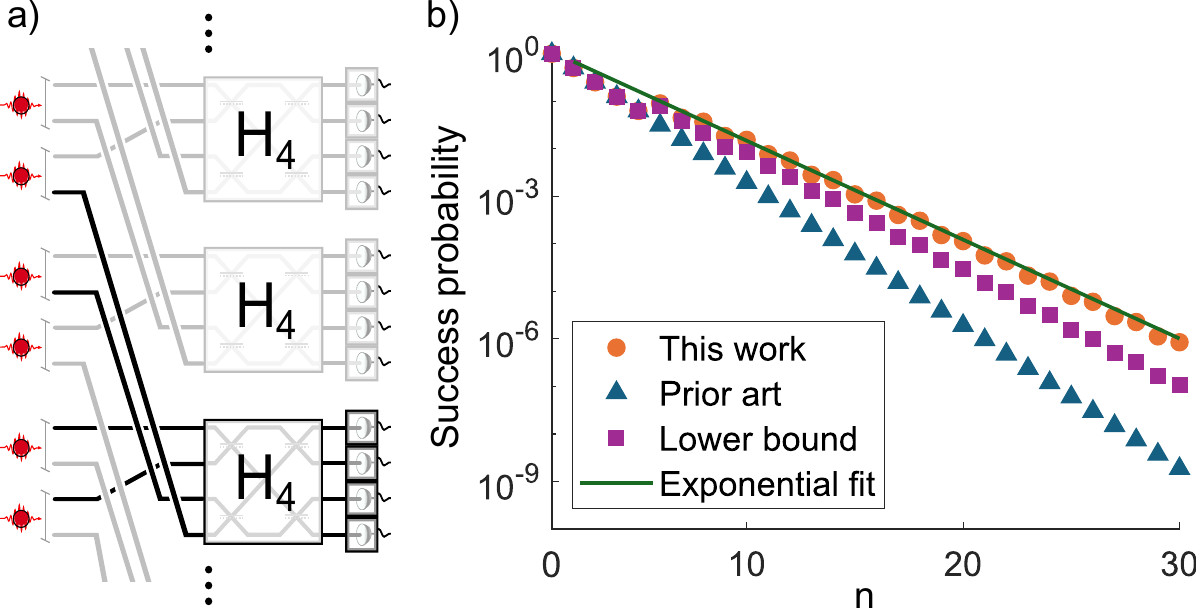}
    \caption{High-efficiency multi-body GHZ measurements. a) Distribution of dual rail qubits over $H_4$ subsystems with periodic boundary conditions. Black rails highlight connectivity to one subsystem. b) Success probability plotted against the size of the projected GHZ-state for the improved scheme presented here (orange disks) and prior-art schemes consisting of cascaded type-I fusions (blue triangles). We additionally display the analytical lower bound form Eq.~\ref{eq main text: lower bound} and an exponential fit to the numerically obtained success probabilities for even values of $n$.}
    \label{fig2: generalization}
\end{figure}

We distribute the dual-rail qubits over the $H_4$ subsystems according to Fig.~(\ref{fig2: generalization}a). The distribution affects both the number of $n$-GHZ states that result in two photons in each subsystem and the probability of successfully projecting onto an $n$-GHZ state, given that there are two photons in each subsystem. The distribution shown in Fig.~(\ref{fig2: generalization}a) yields the most favorable success probability scaling among the distributions we considered.

We analytically derive a lower bound on the success probability of $n-$GHZ projections. Our derivation consists of two parts. First, we count the number of $n-$GHZ states that result in two photons in each subsystem. Second, we count the number of configurations for the subsystems to bunch or anti-bunch such that we retrieve all pairwise Z stabilizers (details can be found in the supplementary material~\cite{SM2}). With this, we arrive at
\begin{equation}
    P>\frac{1}{2^{2N}}\left(L_N\tilde{L}_N\frac{2}{2^N}+(2^N-L_N)\frac{2}{2^N}\right).
    \label{eq main text: lower bound}
\end{equation}
Here, $N$ is the number of $H_4$ subsystems and $n=2 N$. $L_N$ is the Lucas number ($\tilde{L}_N=L_N$ for odd $N$ and $\tilde{L}_N=L_N-2$ for even $N$). $L_N$ is a cousin of the Fibonacci number and can be expressed in terms of the golden ratio $\phi$; $L_N=(\phi^N+(1-\phi)^N)$. We would like to stress that this lower bound scales as $\frac{1}{2^n} (\frac{\phi}{\sqrt{2}})^n\approx 0.572^n$ and thus exponentially outperforms the previously known success probability which scales as $0.5^n$~\cite{PhysRevA.110.032402, Saied2024AQN, bartolucci2021creation}. Numerical evidence suggests that the exact success probability scales significantly better ($\approx0.6192^n$ ) which can be seen in Fig.~(\ref{fig2: generalization}b) up to $n=30$. 

We have focused on the non-adaptive case, but adaptive strategies could ensure projections onto stabilizer states here too. Furthermore, using $H_4$ subsystems restricts our $n$-GHZ projections to an even number of qubits. For an odd number, one can first perform a type-I fusion between two qubits and then apply our proposed projection.

\section{Applications}
\subsection{Resource generation} 
Efficient generation of resource graph-states is a central challenge for photonic quantum technologies, where probabilistic fusion operations typically dominate the resource overhead~\cite{bartolucci2025comparisonschemeshighlyloss, Lee2023graphtheoretical,wein2024minimizingresourceoverheadfusionbased}. Because our four-body GHZ measurement directly entangles four graph-state vertices, it can replace consecutive Bell-measurement-based fusion steps and thereby reduce the expected resource cost.

We first consider graph states generated from a single quantum emitter. Following the deterministic emitter protocol of Refs.~\cite{PhysRevA.111.052604,8l5g-x3b7}, which can generate any graph state with a caterpillar structure, we show in the Supplementary Material~\cite{SM3} that a 14-qubit caterpillar state can be converted into the loss-tolerant cube graph~\cite{PRXQuantum.4.020328} using only one rotated two-body fusion followed by a single four-body GHZ projection. This construction succeeds with probability $9/64 \approx 14\%$, exceeding the success probability obtained from the corresponding sequence of three Bell-type fusion operations $12.5\%$~\cite{PhysRevA.111.052604}.

The same principle reduces the 3-GHZ resource cost of photonic resource-state generation, a useful footprint metric for all-photonic systems~\cite{bartolucci2025comparisonschemeshighlyloss, Lee2023graphtheoretical}. Using the resource-counting framework of Refs.~\cite{bartolucci2025comparisonschemeshighlyloss, Lee2023graphtheoretical}, our four-body GHZ projection reduces the expected cost of the rotated $\{2,2\}$-Shor-encoded 6-ring, a resource state for loss-tolerant photonic quantum computing~\cite{bartolucci2023fusion,wein2024minimizingresourceoverheadfusionbased, Lee2023graphtheoretical}, from $1520$ to approximately $1140$ 3-GHZ states. Moreover, the same resource cost suffices to generate the more loss-tolerant rotated $\{3,2\}$-Shor-encoded 6-ring. Similar reductions are obtained for other graph-state resource states, including the $\{2,2\}$ Shor-encoded 4-star (equivalently the $m=4$ repeater graph), whose expected cost decreases from $256$ to approximately $100$ 3-GHZ states. Details of these constructions are provided in the Supplementary Material~\cite{SM4}. 

\begin{table}[ht]
  \centering
  \begin{tabular}{l|c|c}
    \toprule
                  & Prior scheme & Our scheme \\
    \midrule
    $\{2,2\}$-Shor-encoded 6-ring  & $ 1520$ & $1140$    \\
    $\{3,2\}$-Shor-encoded 6-ring  & $- $ & $1140 $    \\
    $\{2,2\}$-Shor-encoded 4-star  & $256 $ & $100$    \\
    \bottomrule
  \end{tabular}
  \caption{Resource costs in terms of the number of 3-qubit GHZ states to generate three different resource states with application to FBQC. The table compares prior schemes from Ref.~\cite{bartolucci2025comparisonschemeshighlyloss} and our improved scheme utilizing the 4-qubit GHZ measurement.}
  \label{tab:resource_state_gen}
\end{table}
\begin{figure}
    \centering
    \includegraphics[width=1\linewidth]{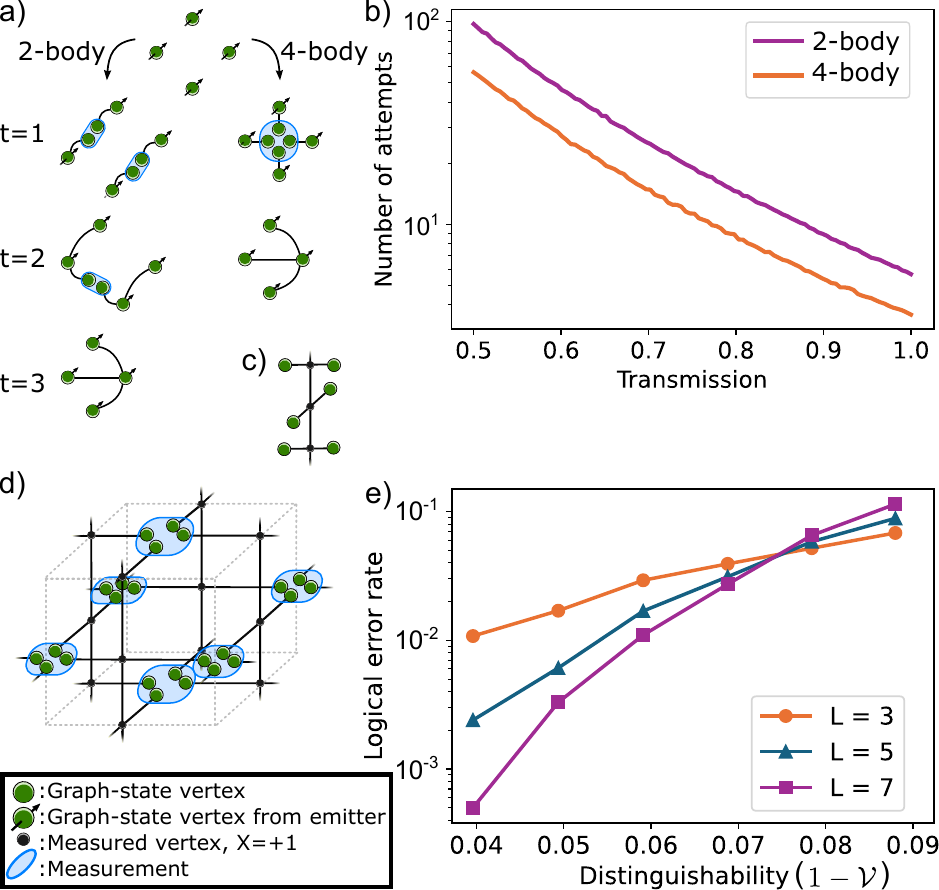}
    \caption{a) Entanglement generation between four spins using either a sequence of two-body fusions or one four-body fusion. b) The average number of entangling attempts using two-body fusions (purple), or using four-body fusions (orange) as a function of the end-to-end photon transmission probability between two spin-photon interfaces. d) A fusion-based decomposition of the Raussendorf lattice using four-body fusions, with the resource state shown in c). e) Distinguishability ($1 - \mathcal{V}$) threshold for the fusion network in d). The logical error rate is obtained by Monte Carlo simulation for different lattice sizes $L$; the threshold is estimated from the intersection of the curves.}
    \label{fig:entang_fbqc}
\end{figure}
\subsection{Entanglement network generation} 
Another promising application of our 4-body GHZ measurement is generating entanglement between non-interacting spin-photon interfaces, an important building block for fault-tolerant photonic quantum computing and multi-party quantum communication networks~\cite{Borregaard2019, Ruskuc2025, Knaut2024}.
We consider four spin-photon interfaces that generate one photon entangled with its spin and send it to the measurement circuit (see Fig.~(\ref{fig:entang_fbqc}a)). 
A successful 4-body measurement projects all spins onto a four-qubit GHZ state, while failed projections result in a new attempt, and photon loss leads to reinitialization of the spins and a new attempt. 
Instead, using two-body fusions~\cite{PhysRevLett.95.010501}, we first have to entangle the spins pairwise through photon emission and two-body fusions. Only then can we entangle one spin of each pair to realize entanglement between all four parties.
Again, photon loss or fusion failure lead to reinitialization and/or a new attempt, however, if either photon is lost in the second entangling step, we have to repeat the first entangling operations.
As shown in Fig.~(\ref{fig:entang_fbqc}b), the 4-body fusion reduces the average number of attempts by a factor of $1.7$.
\subsection{Fault-tolerant computation}
Another promising application of our adaptive 4-body GHZ measurement is fusion-based quantum computing (FBQC)~\cite {bartolucci2023fusion}.
FBQC is achieved through destructive and probabilistic entangling gates between resource states arranged in a fusion network defining a fault-tolerant code.
The parity outcomes drive the computation and provide syndrome data; see Ref.~\cite {bartolucci2023fusion} for details on FBQC.
Importantly, we require the projective parity measurements to be stabilizer measurements; thus, we require the adaptive fusion circuit that removes the $W$-state and $D$-state projections.
In Fig.~\ref{fig:entang_fbqc} c)-d), a promising decomposition of the Raussendorf lattice~\cite{Raussendorf_2007} into infinite lines is shown. These resource states are compatible with quantum emitters, as these can be generated deterministically from a single emitter.
Leveraging the 4-body GHZ measurement removes the need for ancilla resource states as used in Ref.~\cite{chen2025fusionbasedimplementationqldpccodes}, lowering the required emitter count by a factor of 2.
Using a repeat-until-success strategy and the methods described in Refs.~\cite{PRXQuantum.6.020304, QuandelaRUS}, we find a distinguishability threshold of $1 - \mathcal{V} = 7.36\%$, shown in Fig.~\ref{fig:entang_fbqc} e), where $\mathcal{V}$ is the Hong-Ou-Mandel dip visibility~\cite{PRXQuantum.6.020304}.
This is an $84\%$ increase in the tolerable distinguishability compared with the $4\%$ threshold reported for the two-body-fusion RUS architecture of Ref.~\cite{PRXQuantum.6.020304}.
\section{Conclusion}

We have introduced a high-efficiency linear-optical GHZ measurement that generalizes photonic Bell measurement from two to many photons. For the four-photon case, our protocol increases the success probability from $1/8$ to $9/32$, while adaptive variants ensure that unsuccessful events still project onto useful stabilizer states. More generally, we show that this approach extends naturally to larger GHZ measurements, yielding an exponentially higher success probability than previously known schemes, although the overall success probability still drops significantly with the number of photons.

Beyond the measurement itself, our results demonstrate that multiphoton interference can be exploited as a practical resource for photonic quantum technologies. We show that four-body GHZ measurements reduce the overhead of photonic graph-state generation, improve the efficiency of entanglement generation between remote spin-photon interfaces, and enable more resource-efficient fusion-based quantum computing architectures that can benefit from an improved distinguishability threshold.

More broadly, our work illustrates that photonic quantum technologies need not be built exclusively from Bell measurements. Instead, native multiphoton interference can serve as a powerful primitive for generating large amounts of entanglement in a single probabilistic operation. Since ancillary photonic resource states can boost the success probability of Bell measurements, similar strategies are likely to further enhance the high-efficiency GHZ measurements introduced here. We expect that this perspective will motivate the search for many-body building blocks beyond linear optics and inspire new architectures that more closely exploit the capabilities of their underlying physical platforms.\\

\textbf{Note added:} During the preparation of this manuscript, we became aware of a patent application~\cite{WO2025212141}, which proposes a ``four-way fusion" scheme equivalent to the scheme we present in Fig.~(\ref{fig1: GHZ projection}). Our work extends this building block by 1) introducing adaptivity to ensure projections on stabilizer states, 2) generalizing the four-way fusion to n-GHZ measurements with exponentially improved success probability, and 3) presenting examples of practically relevant applications of the n-GHZ projection schemes, which outperform prior-art methods. 

\section{Ackowledgements}
We thank Jelmer Renema for useful discussions. 
We acknowledge financial support from Danmarks Innovationsfond (IFD1003402609, FTQP).
S.P. acknowledges funding from VILLUM FONDEN (No. VIL60743 and No. VIL78724), the European Research Council (No. 101221875), Danmarks Innovationsfond
research grant (No. 4356-00009B) and from the NNF Quantum Computing Programme. S.H. acknowledges financial support through the PhotonDelta National Growth Fund programme.

\section{Data availability}
The data that support the findings of
this article are openly available \cite{PT/PLFNIN_2026}.



\bibliographystyle{apsrev}

\bibliography{biblio.bib}

@data{PT/PLFNIN_2026,
author = {van den Hoven, Stefan Nicolaas and Pettersson, Love Alexander Mandla and Sørensen, Anders Søndberg and Paesani, Stefano},
publisher = {DANS Data Station Physical and Technical Sciences},
title = {{Replication Data for: ``High-efficiency photonic GHZ measurements with linear optics''}},
year = {2026},
version = {DRAFT VERSION},
doi = {10.17026/PT/PLFNIN},
url = {https://doi.org/10.17026/PT/PLFNIN}
}

@misc{SM1,
  note = {See Supplemental Material for a illustrations of adaptive schemes that ensure projections onto stabilizer states upon failure.}
}

@misc{SM2,
  note = {See Supplemental Material for details on the analytical derivation of lower bound on success probability of n-GHZ projections.}
}

@misc{SM3,
  note = {See Supplemental Material for details on the generation of a loss-tolerant cube graph from a 14-qubit caterpillar state using only one rotated two-body fusion followed by a single four-body GHZ projection.}
}

@misc{SM4,
  note = {See Supplemental Material for details on the preparation of the $\{2,2\}$- and the $\{3,2\}$-Shor-encoded 6-ring as well as the $\{2,2\}$-Shor-encoded 4-star.}
}

@misc{gottesman1998heisenbergrepresentationquantumcomputers,
      title={The Heisenberg Representation of Quantum Computers}, 
      author={Daniel Gottesman},
      year={1998},
      eprint={quant-ph/9807006},
      archivePrefix={arXiv},
      primaryClass={quant-ph},
      url={https://arxiv.org/abs/quant-ph/9807006}, 
}

@article{PhysRevA.68.032310,
  title = {Measurement-induced nonlinearity in linear optics},
  author = {Scheel, Stefan and Nemoto, Kae and Munro, William J. and Knight, Peter L.},
  journal = {Phys. Rev. A},
  volume = {68},
  issue = {3},
  pages = {032310},
  numpages = {13},
  year = {2003},
  month = {Sep},
  publisher = {American Physical Society},
  doi = {10.1103/PhysRevA.68.032310},
  url = {https://link.aps.org/doi/10.1103/PhysRevA.68.032310}
}

@article{Knaut2024,
  title = {Entanglement of nanophotonic quantum memory nodes in a telecom network},
  volume = {629},
  ISSN = {1476-4687},
  url = {http://dx.doi.org/10.1038/s41586-024-07252-z},
  DOI = {10.1038/s41586-024-07252-z},
  number = {8012},
  journal = {Nature},
  publisher = {Springer Science and Business Media LLC},
  author = {Knaut,  C. M. and Suleymanzade,  A. and Wei,  Y.-C. and Assumpcao,  D. R. and Stas,  P.-J. and Huan,  Y. Q. and Machielse,  B. and Knall,  E. N. and Sutula,  M. and Baranes,  G. and Sinclair,  N. and De-Eknamkul,  C. and Levonian,  D. S. and Bhaskar,  M. K. and Park,  H. and Lončar,  M. and Lukin,  M. D.},
  year = {2024},
  month = May,
  pages = {573–578}
}

@article{Borregaard2019,
  title = {Quantum Networks with Deterministic Spin–Photon Interfaces},
  volume = {2},
  ISSN = {2511-9044},
  url = {http://dx.doi.org/10.1002/qute.201800091},
  DOI = {10.1002/qute.201800091},
  number = {5-6},
  journal = {Advanced Quantum Technologies},
  publisher = {Wiley},
  author = {Borregaard,  Johannes and Sørensen,  Anders Søndberg and Lodahl,  Peter},
  year = {2019},
  month = Apr 
}

@article{Ruskuc2025,
  title = {Multiplexed entanglement of multi-emitter quantum network nodes},
  volume = {639},
  ISSN = {1476-4687},
  url = {http://dx.doi.org/10.1038/s41586-024-08537-z},
  DOI = {10.1038/s41586-024-08537-z},
  number = {8053},
  journal = {Nature},
  publisher = {Springer Science and Business Media LLC},
  author = {Ruskuc,  A. and Wu,  C.-J. and Green,  E. and Hermans,  S. L. N. and Pajak,  W. and Choi,  J. and Faraon,  A.},
  year = {2025},
  month = Feb,
  pages = {54–59}
}

@article{Oh:25,
author = {Changhun Oh},
journal = {Curr. Opt. Photon.},
number = {1},
pages = {1--18},
publisher = {Optica Publishing Group},
title = {Recent Theoretical and Experimental Progress on Boson Sampling},
volume = {9},
month = {Feb},
year = {2025},
url = {https://opg.optica.org/copp/abstract.cfm?URI=copp-9-1-1},
}

@misc{schadow2026certificationlinearopticalquantum,
      title={Certification of linear optical quantum state preparation}, 
      author={Riko Schadow and Naomi Spier and Stefan N. van den Hoven and Malaquias Correa Anguita and Redlef B. G. Braamhaar and Sara Marzban and Jens Eisert and Jelmer J. Renema and Nathan Walk},
      year={2026},
      eprint={2602.12269},
      archivePrefix={arXiv},
      primaryClass={quant-ph},
      url={https://arxiv.org/abs/2602.12269}, 
}

@article{hoven2026quantumadvantagesinglephotonstate,
  title = {Quantum Advantage for Single-Photon State Characterization},
  author = {van den Hoven, S. N. and Anguita, M. C. and Marzban, S. and Renema, J. J.},
  journal = {Phys. Rev. Lett.},
  volume = {137},
  issue = {7},
  pages = {073604},
  numpages = {6},
  year = {2026},
  month = {Aug},
  publisher = {American Physical Society},
  doi = {10.1103/g28d-jzgj},
  url = {https://link.aps.org/doi/10.1103/g28d-jzgj}
}

@misc{bartolucci2025comparisonschemeshighlyloss,
      title={Comparison of schemes for highly loss tolerant photonic fusion based quantum computing}, 
      author={Sara Bartolucci and Tom Bell and Hector Bombin and Patrick Birchall and Jacob Bulmer and Christopher Dawson and Terry Farrelly and Samuel Gartenstein and Mercedes Gimeno-Segovia and Daniel Litinski and Yehua Liu and Robert Knegjens and Naomi Nickerson and Andrea Olivo and Mihir Pant and Ashlesha Patil and Sam Roberts and Terry Rudolph and Chris Sparrow and David Tuckett and Andrzej Veitia},
      year={2025},
      eprint={2506.11975},
      archivePrefix={arXiv},
      primaryClass={quant-ph},
      url={https://arxiv.org/abs/2506.11975}, 
}

@article{Lee2023graphtheoretical,
  doi = {10.22331/q-2023-12-20-1212},
  url = {https://doi.org/10.22331/q-2023-12-20-1212},
  title = {Graph-theoretical optimization of fusion-based graph state generation},
  author = {Lee, Seok-Hyung and Jeong, Hyunseok},
  journal = {{Quantum}},
  issn = {2521-327X},
  publisher = {{Verein zur F{\"{o}}rderung des Open Access Publizierens in den Quantenwissenschaften}},
  volume = {7},
  pages = {1212},
  month = dec,
  year = {2023}
}

@article{PRXQuantum.4.020328,
  title = {Optimizing Graph Codes for Measurement-Based Loss Tolerance},
  author = {Bell, Thomas J. and Pettersson, Love A. and Paesani, Stefano},
  journal = {PRX Quantum},
  volume = {4},
  issue = {2},
  pages = {020328},
  numpages = {20},
  year = {2023},
  month = {May},
  publisher = {American Physical Society},
  doi = {10.1103/PRXQuantum.4.020328},
  url = {https://link.aps.org/doi/10.1103/PRXQuantum.4.020328}
}

@article{8l5g-x3b7,
  title = {Minimizing Resource Overhead in Fusion-Based Quantum Computation Using Hybrid Spin-Photon Devices},
  author = {Wein, Stephen C. and Goubault de Brugi\`ere, Timoth\'ee and Music, Luka and Senellart, Pascale and Bourdoncle, Boris and Mansfield, Shane},
  journal = {PRX Quantum},
  volume = {6},
  issue = {4},
  pages = {040362},
  numpages = {26},
  year = {2025},
  month = {Dec},
  publisher = {American Physical Society},
  doi = {10.1103/8l5g-x3b7},
  url = {https://link.aps.org/doi/10.1103/8l5g-x3b7}
}

@article{PhysRevA.111.052604,
  title = {Generating graph states with a single quantum emitter and the minimum number of fusions},
  author = {L\"obl, Matthias C. and Pettersson, Love A. and Jena, Andrew and Dellantonio, Luca and Paesani, Stefano and S\o{}rensen, Anders S.},
  journal = {Phys. Rev. A},
  volume = {111},
  issue = {5},
  pages = {052604},
  numpages = {16},
  year = {2025},
  month = {May},
  publisher = {American Physical Society},
  doi = {10.1103/PhysRevA.111.052604},
  url = {https://link.aps.org/doi/10.1103/PhysRevA.111.052604}
}

@article{Calsamiglia2001,
  author    = {John Calsamiglia and Norbert L{\"u}tkenhaus},
  title     = {Maximum Efficiency of a Linear-Optical Bell-State Analyzer},
  journal   = {Applied Physics B},
  year      = {2001},
  volume    = {72},
  number    = {1},
  pages     = {67--71},
  doi       = {10.1007/s003400000484},
  url       = {https://doi.org/10.1007/s003400000484}
}

@article{Kimble2008,
  author  = {H. J. Kimble},
  title   = {The Quantum Internet},
  journal = {Nature},
  year    = {2008},
  volume  = {453},
  number  = {7198},
  pages   = {1023--1030},
  doi     = {10.1038/nature07127},
  url     = {https://doi.org/10.1038/nature07127}
}

@article{
doi:10.1126/science.aam9288,
author = {Stephanie Wehner  and David Elkouss  and Ronald Hanson },
title = {Quantum internet: A vision for the road ahead},
journal = {Science},
volume = {362},
number = {6412},
pages = {eaam9288},
year = {2018},
doi = {10.1126/science.aam9288},
URL = {https://www.science.org/doi/abs/10.1126/science.aam9288},
eprint = {https://www.science.org/doi/pdf/10.1126/science.aam9288}}

@article{RevModPhys.95.045006,
  title = {Quantum repeaters: From quantum networks to the quantum internet},
  author = {Azuma, Koji and Economou, Sophia E. and Elkouss, David and Hilaire, Paul and Jiang, Liang and Lo, Hoi-Kwong and Tzitrin, Ilan},
  journal = {Rev. Mod. Phys.},
  volume = {95},
  issue = {4},
  pages = {045006},
  numpages = {66},
  year = {2023},
  month = {Dec},
  publisher = {American Physical Society},
  doi = {10.1103/RevModPhys.95.045006},
  url = {https://link.aps.org/doi/10.1103/RevModPhys.95.045006}
}

@article{PhysRevLett.81.5932,
  title = {Quantum Repeaters: The Role of Imperfect Local Operations in Quantum Communication},
  author = {Briegel, H.-J. and D\"ur, W. and Cirac, J. I. and Zoller, P.},
  journal = {Phys. Rev. Lett.},
  volume = {81},
  issue = {26},
  pages = {5932--5935},
  numpages = {0},
  year = {1998},
  month = {Dec},
  publisher = {American Physical Society},
  doi = {10.1103/PhysRevLett.81.5932},
  url = {https://link.aps.org/doi/10.1103/PhysRevLett.81.5932}
}

@misc{WO2025212141,
  author       = {Bartolucci, Sara and Pant, Mihir and Rudolph, Terrence and Sparrow, Christopher},
  title        = {Multi-way fusion circuits for photonic qubits},
  howpublished = {International Patent Application WO 2025/212141 A1},
  year         = {2025},
  month        = nov,
  note         = {Published under the Patent Cooperation Treaty (PCT)},
  doi          = {},
  url          = {https://patentscope.wipo.int/search/en/WO2025212141}
}

@misc{ostmann2025nonlinearphotonicarchitecturefaulttolerant,
      title={Nonlinear photonic architecture for fault-tolerant quantum computing}, 
      author={Maike Ostmann and Joshua Nunn and Alex E. Jones},
      year={2025},
      eprint={2510.06890},
      archivePrefix={arXiv},
      primaryClass={quant-ph},
      url={https://arxiv.org/abs/2510.06890}, 
}

@misc{wein2024minimizingresourceoverheadfusionbased,
      title={Minimizing resource overhead in fusion-based quantum computation using hybrid spin-photon devices}, 
      author={Stephen C. Wein and Timothée Goubault de Brugière and Luka Music and Pascale Senellart and Boris Bourdoncle and Shane Mansfield},
      year={2024},
      eprint={2412.08611},
      archivePrefix={arXiv},
      primaryClass={quant-ph},
      url={https://arxiv.org/abs/2412.08611}, 
}

@article{chen2025fusionbasedimplementationqldpccodes,
  title = {Fusion-based implementation of qLDPC codes with quantum emitters},
  volume = {12},
  ISSN = {2056-6387},
  url = {http://dx.doi.org/10.1038/s41534-026-01233-y},
  DOI = {10.1038/s41534-026-01233-y},
  number = {1},
  journal = {npj Quantum Information},
  publisher = {Springer Science and Business Media LLC},
  author = {Chen,  Susan X. and L\"{o}bl,  Matthias C. and Chan,  Ming Lai and Sørensen,  Anders S. and Paesani,  Stefano},
  year = {2026},
  month = Apr 
}

@article{PRXQuantum.6.020304,
  title = {Tailoring Fusion-Based Photonic Quantum Computing Schemes to Quantum Emitters},
  author = {Chan, Ming Lai and Bell, Thomas J. and Pettersson, Love A. and Chen, Susan X. and Yard, Patrick and S\o{}rensen, Anders S. and Paesani, Stefano},
  journal = {PRX Quantum},
  volume = {6},
  issue = {2},
  pages = {020304},
  numpages = {18},
  year = {2025},
  month = {Apr},
  publisher = {American Physical Society},
  doi = {10.1103/PRXQuantum.6.020304},
  url = {https://link.aps.org/doi/10.1103/PRXQuantum.6.020304}
}

@article{Raussendorf_2007,
doi = {10.1088/1367-2630/9/6/199},
url = {https://doi.org/10.1088/1367-2630/9/6/199},
year = {2007},
month = {jun},
publisher = {},
volume = {9},
number = {6},
pages = {199},
author = {Raussendorf, R and Harrington, J and Goyal, K},
title = {Topological fault-tolerance in cluster state quantum computation},
journal = {New Journal of Physics}
}

@misc{QuandelaRUS,
      title={Enhanced Fault-tolerance in Photonic Quantum Computing: Comparing the Honeycomb Floquet Code and the Surface Code in Tailored Architecture}, 
      author={Théo Dessertaine and Boris Bourdoncle and Aurélie Denys and Grégoire de Gliniasty and Pierre Colonna d'Istria and Gerard Valentí-Rojas and Shane Mansfield and Paul Hilaire},
      year={2026},
      eprint={2410.07065},
      archivePrefix={arXiv},
      primaryClass={quant-ph},
      url={https://arxiv.org/abs/2410.07065}, 
}

@misc{chan2025practicalblueprintlowdepthphotonic,
      title={Practical blueprint for low-depth photonic quantum computing with quantum dots}, 
      author={Ming Lai Chan and Aliki Anna Capatos and Peter Lodahl and Anders Søndberg Sørensen and Stefano Paesani},
      year={2025},
      eprint={2507.16152},
      archivePrefix={arXiv},
      primaryClass={quant-ph},
      url={https://arxiv.org/abs/2507.16152}, 
}

@article{PhysRevA.110.032402,
  title = {Flexible entangled-state generation in linear optics},
  author = {Pankovich, Brendan and Neville, Alex and Kan, Angus and Omkar, Srikrishna and Wan, Kwok Ho and Br\'adler, Kamil},
  journal = {Phys. Rev. A},
  volume = {110},
  issue = {3},
  pages = {032402},
  numpages = {17},
  year = {2024},
  month = {Sep},
  publisher = {American Physical Society},
  doi = {10.1103/PhysRevA.110.032402},
  url = {https://link.aps.org/doi/10.1103/PhysRevA.110.032402}
}

@article{PhysRevLett.95.010501,
  title = {Resource-Efficient Linear Optical Quantum Computation},
  author = {Browne, Daniel E. and Rudolph, Terry},
  journal = {Phys. Rev. Lett.},
  volume = {95},
  issue = {1},
  pages = {010501},
  numpages = {4},
  year = {2005},
  month = {Jun},
  publisher = {American Physical Society},
  doi = {10.1103/PhysRevLett.95.010501},
  url = {https://link.aps.org/doi/10.1103/PhysRevLett.95.010501}
}

@article{PhysRevLett.113.140403,
  title = {$3/4$-Efficient Bell Measurement with Passive Linear Optics and Unentangled Ancillae: boosted fusion},
  author = {Ewert, Fabian and van Loock, Peter},
  journal = {Phys. Rev. Lett.},
  volume = {113},
  issue = {14},
  pages = {140403},
  numpages = {5},
  year = {2014},
  month = {Sep},
  publisher = {American Physical Society},
  doi = {10.1103/PhysRevLett.113.140403},
  url = {https://link.aps.org/doi/10.1103/PhysRevLett.113.140403}
}

@article{PhysRevA.84.042331,
  title = {Arbitrarily complete Bell-state measurement using only linear optical elements, boosted fusion},
  author = {Grice, W. P.},
  journal = {Phys. Rev. A},
  volume = {84},
  issue = {4},
  pages = {042331},
  numpages = {6},
  year = {2011},
  month = {Oct},
  publisher = {American Physical Society},
  doi = {10.1103/PhysRevA.84.042331},
  url = {https://link.aps.org/doi/10.1103/PhysRevA.84.042331}
}

@phdthesis{gottesman1997stabilizer,
  title={Stabilizer codes and quantum error correction},
  author={Gottesman, Daniel},
  year={1997},
  publisher={California Institute of Technology}
}

@article{bartolucci2023fusion,
  title={Fusion-based quantum computation},
  author={Bartolucci, Sara and Birchall, Patrick and Bombin, Hector and Cable, Hugo and Dawson, Chris and Gimeno-Segovia, Mercedes and Johnston, Eric and Kieling, Konrad and Nickerson, Naomi and Pant, Mihir and others},
  journal={Nat. Commun.},
  volume={14},
  number={1},
  pages={912},
  year={2023},
  publisher={Nature Publishing Group UK London}
}

@article{renema2018efficient,
  title={Efficient classical algorithm for boson sampling with partially distinguishable photons},
  author={Renema, Jelmer J and Menssen, Adrian and Clements, William R and Triginer, Gil and Kolthammer, William S and Walmsley, Ian A},
  journal={Phys. Rev. Lett.},
  volume={120},
  number={22},
  pages={220502},
  year={2018},
  publisher={APS}
}

@inproceedings{Aaronson_Arkhipov_BS,
author = {Aaronson, Scott and Arkhipov, Alex},
title = {The computational complexity of linear optics},
year = {2011},
isbn = {9781450306911},
publisher = {Association for Computing Machinery},
address = {New York, NY, USA},
booktitle = {Proceedings of the Forty-Third Annual ACM Symposium on Theory of Computing},
pages = {333–342},
numpages = {10},
location = {San Jose, California, USA},
series = {STOC '11}
}

@misc{Saied2024AQN,
  title={Advancing Quantum Networking: Some Tools and Protocols for Ideal and Noisy Photonic Systems},
  author={Jason Saied and Jeffrey Marshall and Namit Anand and Shon Grabbe and Eleanor G. Rieffel},
  eprint={2403.02515},
  archivePrefix={arXiv},
  PrimaryClass={quant-ph},
  year={2024}
}

@article{knill2001scheme,
  title={A scheme for efficient quantum computation with linear optics},
  author={Knill, Emanuel and Laflamme, Raymond and Milburn, Gerald J},
  journal={nature},
  volume={409},
  number={6816},
  pages={46--52},
  year={2001},
  publisher={Nature Publishing Group UK London}
}

@misc{bartolucci2021creation,
  title={Creation of entangled photonic states using linear optics},
  author={Bartolucci, Sara and Birchall, Patrick M and Gimeno-Segovia, Mercedes and Johnston, Eric and Kieling, Konrad and Pant, Mihir and Rudolph, Terry and Smith, Jake and Sparrow, Chris and Vidrighin, Mihai D},
  eprint={2106.13825},
  archivePrefix={arXiv},
  primaryClass={quant-ph},
  year={2021}
}

@misc{somhorst2024photon,
      title={Photon distillation schemes with reduced resource costs based on multiphoton Fourier interference}, 
      author={F. H. B. Somhorst and B. K. Sauër and S. N. van den Hoven and J. J. Renema},
      year={2024},
      eprint={2404.14262},
      archivePrefix={arXiv},
      primaryClass={quant-ph}
}


\newpage 
\clearpage

\pagenumbering{arabic}
\appendix
\renewcommand{\thesection}
{\Alph{section}}
\renewcommand{\thefigure}{\thesection\arabic{figure}}
\setcounter{figure}{0} 
\renewcommand{\thetable}{\thesection\arabic{table}}
\setcounter{table}{0} 


\onecolumngrid
\section{Appendix}
\section{Additional figures}

\begin{figure}[h!]
    \centering
    \includegraphics[width=1.0\linewidth]{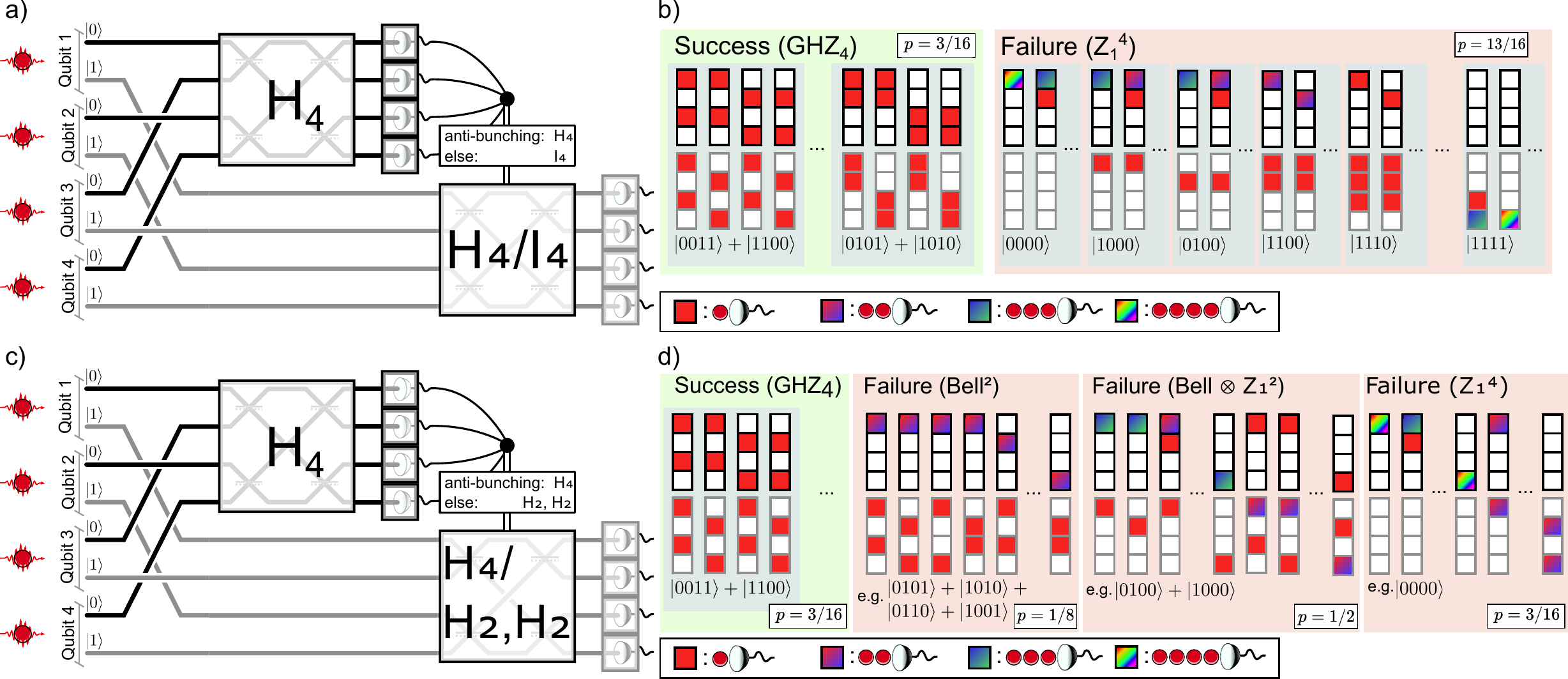}
    \caption{Adaptive four-body GHZ measurements. a) Linear-optical network implementing an adaptive GHZ measurement in which the interferometer in the second subsystem is chosen based on the detection outcome of the first subsystem: $H_4$ is used when the first subsystem detects two photons in an anti-bunched configuration and $I_4$ otherwise. Double wires indicate the classical feed-forward of the detection outcome. b) Representative set of corresponding heralding patterns and projections for this adaptive scheme. Here, anti-bunching denotes detection of two photons in each of four distinct output modes; in the four-mode notation, the six anti-bunched patterns are $(0011),(1100),(0110),(1001),(1010),(0101)$. The scheme projects onto a $4$-GHZ state with probability $3/16$, while all other outcomes project the four qubits in the computational ($Z$) basis, with probability $13/16$. c) Alternative adaptive scheme in which the second subsystem is switched between $H_4$ and two parallel $H_2$ interferometers according to the same feed-forward rule. d) Representative set of corresponding heralding patterns and projections. The scheme succeeds with probability $3/16$; failure outcomes project either onto two separable Bell states, denoted $\mathrm{Bell}^2$, with probability $1/8$, onto one Bell pair and two qubits in the $Z$ basis, denoted $\mathrm{Bell}\otimes Z_1^2$, with probability $1/2$, or onto the four-qubit $Z$ basis with probability $3/16$. In the $\mathrm{Bell}\otimes Z_1^2$ outcomes, the Bell pair involves either qubits $1,2$ or qubits $3,4$, each with probability $1/4$.}
    \label{fig: adaptvie GHZ projection}
\end{figure}

\clearpage

\begin{figure}
    \centering
    \includegraphics[width=1\linewidth]{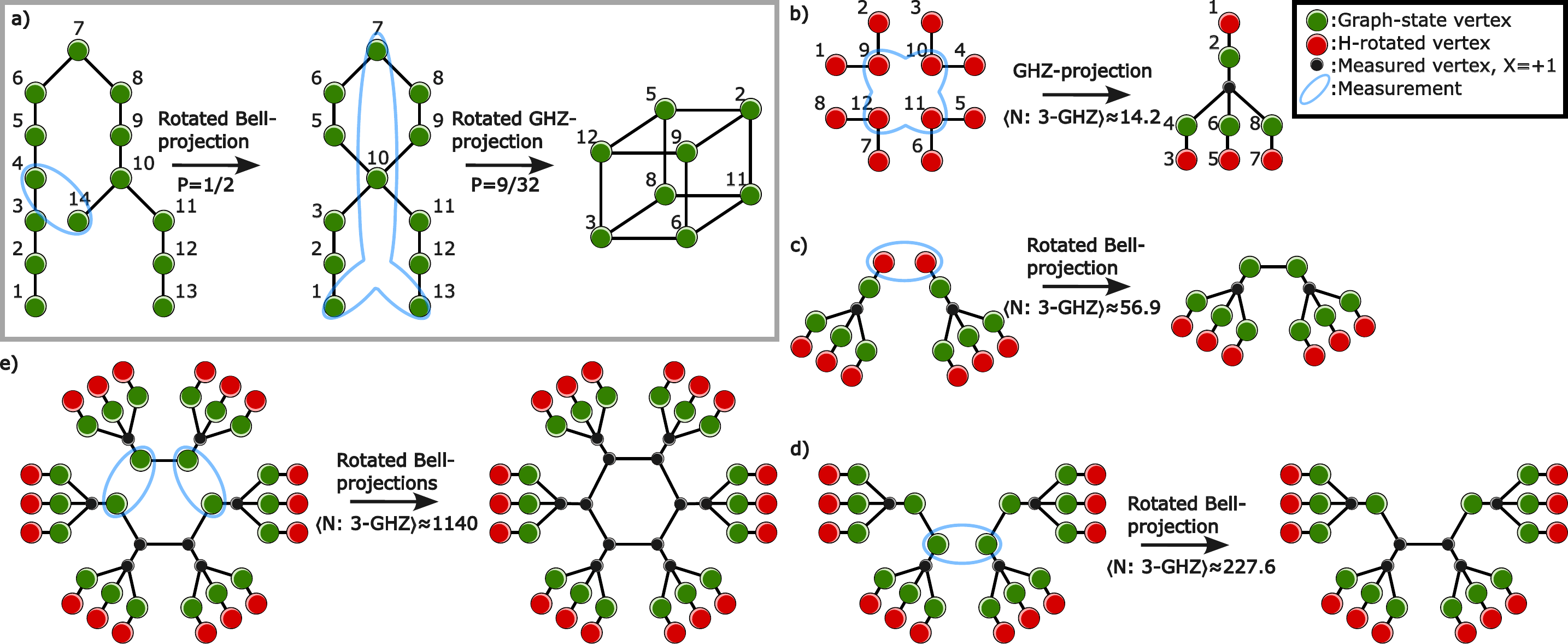}
    \caption{Resource-state generation enabled by high-efficiency GHZ measurements. a) Generation of the cube graph from a caterpillar graph state using a rotated two-body Bell projection ($\{X_4Z_{14}, Z_{4}X_{14}\}$) followed by a rotated four-body GHZ projection ($\{X_{1}X_{7}Z_{10}X_{13}, Z_{1}Z_{7}, Z_{7}X_{10}, X_{10}Z_{13}\}$), resulting in an overall success probability of $\approx 14\%$. b)–e) Step-by-step generation of the rotated $\{3,2\}$-Shor-encoded 6-ring from 3-GHZ states using high-efficiency GHZ projections and rotated Bell projections ($\{XZ, ZX\}$). The values $\langle N: 3\text{-GHZ}\rangle$ denote the cumulative expected number of 3-GHZ states required for each intermediate construction. Graph-state vertices are shown in green, Hadamard-rotated vertices in red, and vertices destructively measured in the $X$ basis with outcome $+1$ in black. Blue loops indicate the vertices involved in the corresponding projection.}
    \label{fig:resource_state_gen32}
\end{figure}

\begin{figure}
    \centering
    \includegraphics[width=1\linewidth]{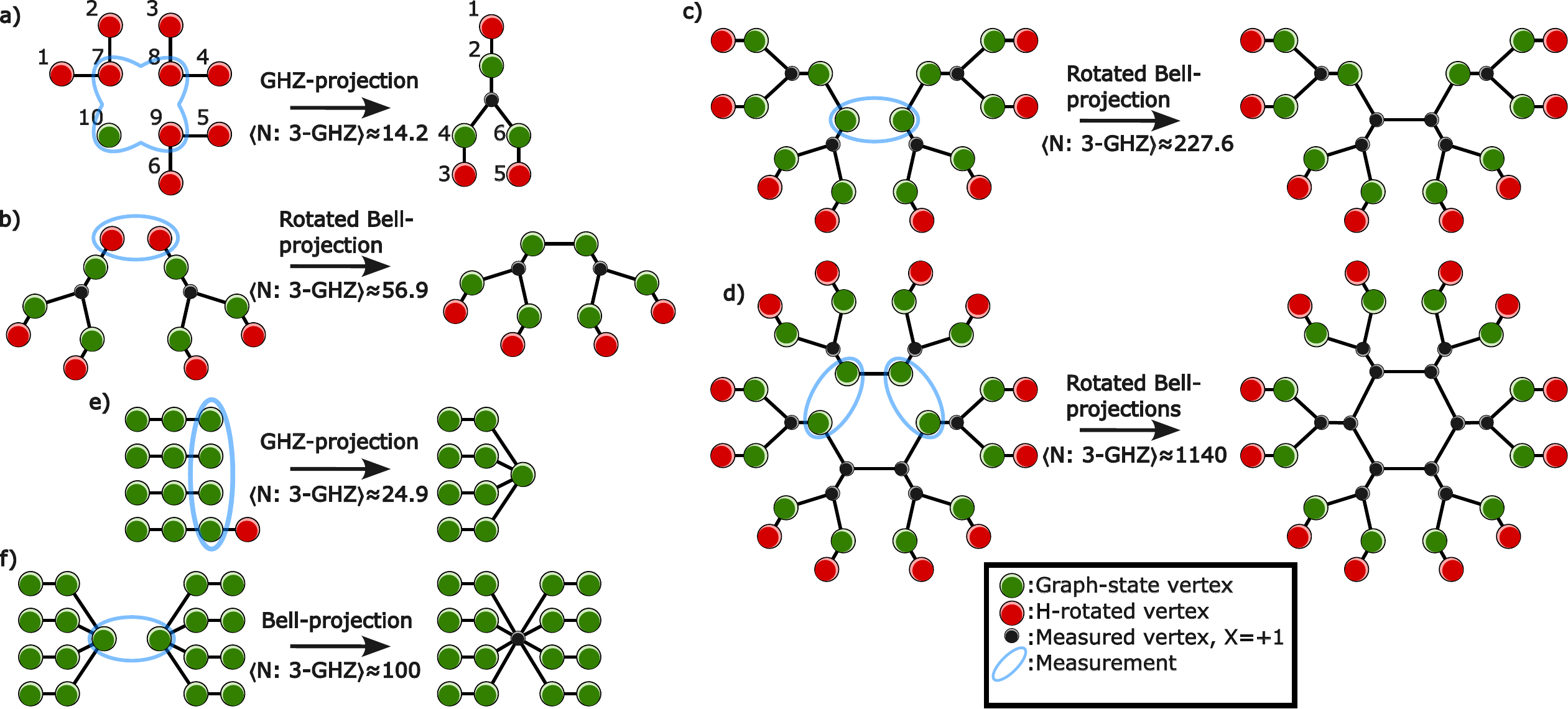}
    \caption{Resource-state generation enabled by high-efficiency GHZ measurements. a)–d) Step-by-step generation of the rotated $\{2,2\}$-Shor-encoded 6-ring from 3-GHZ states using high-efficiency GHZ projections and rotated Bell projections ($\{XZ, ZX\}$). e)-f) Step-by-step generation of the rotated $\{2,2\}$-Shor-encoded 4-star from 3-GHZ states using high-efficiency GHZ projections and Bell projections ($\{XX, ZZ\}$). The values $\langle \text{N: 3-GHZ}\rangle$ denote the cumulative expected number of 3-GHZ states required for each intermediate construction. Graph-state vertices are shown in green, Hadamard-rotated vertices in red, and vertices destructively measured in the $X$ basis with outcome $+1$ in black. Blue loops indicate the vertices involved in the corresponding projection.}
    \label{fig:resource_state_gen22}
\end{figure}

\clearpage

\twocolumngrid

\section{Analytical derivation of lower bound on success probability of n-GHZ projections}

\begin{figure}[h!]
    \centering
    \includegraphics[width=0.9\linewidth]{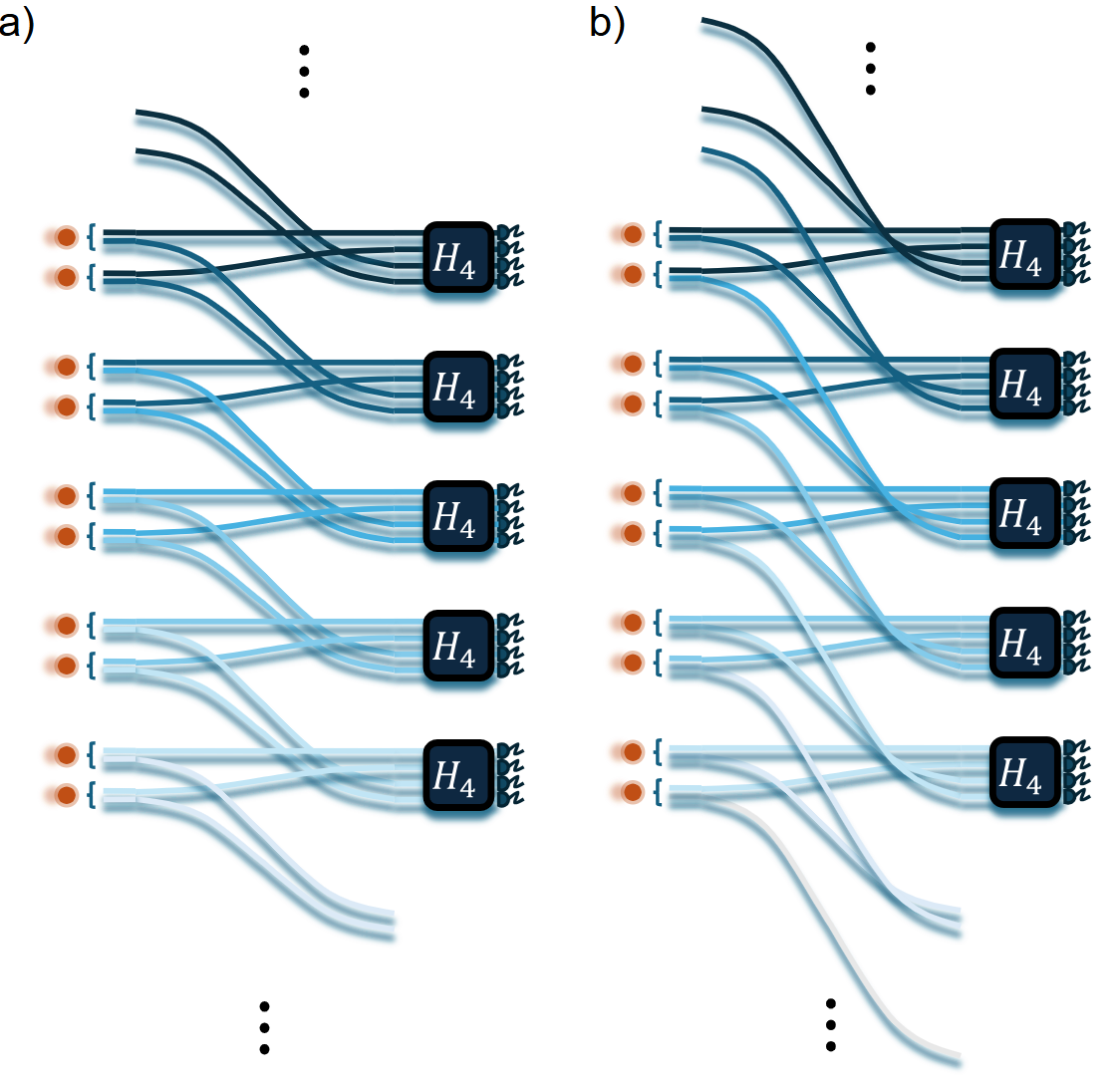}
    \caption{Two different ways to connect the dual-rail qubits to the Hadamard subsystems.}
    \label{fig: connections}
\end{figure}

In this section, we will derive a lower bound on the success probability of the n-GHZ projections as presented in the main text. We do this for two different methods to connect the dual rails of the input qubits to the $H_4$ subsystems. The different connections will result in a different number of n-GHZ states that result in two photons entering each subsystem. The different connections will also result into the learning of different sets of Z generators whenever a subsystem bunches. Therefore, different configurations of subsystems that are allowed to bunch to successfully retrieve the n-GHZ projection will differ. The two different connections are presented in Fig. (\ref{fig: connections}).

A convenient way to describe these different connections, is in terms of the following generating functions:
\begin{equation}
    g_1=\prod_{i=1}^N(u_i+u_{\mathrm{Mod}[i+1,N,1]})^2,
    \label{eq: g1}
\end{equation}
\begin{equation}
    g_2=\prod_{i=1}^N(u_i+u_{\mathrm{Mod}[i+1,N,1]})(u_{i}+u_{\mathrm{Mod}[i+2,N,1]}).
    \label{eq: g2}
\end{equation}
Here, $N$ is the total number of subsystems and thus equal to $n/2$ and Mod$[N+k,N,1]=k$. In the remainder of the text, we will not bother to write Mod[..] for visual clarity, but we keep in the back of our minds that subsystems are periodically attached to each other. Each factor in the generating function represents a dual rail qubit. The fact that there are two terms in each factor indicates that each qubit is connected to two subsystems, the first term represents the zero-rail, the second term represents the one-rail. By expanding the brackets and inspecting the coefficients of the polynomial, we count the number of computational basis states that result in a specific number of photons entering certain subsystems. For example: the coefficient of $\prod_{i=1}^Nu_i^2$ counts the number of computational basis states that will result in detecting two photons in each subsystem. Since every n-GHZ state is a superposition of two computational basis states that are each others complement (i.e. one term is equal to the other where every bit is flipped), and each subsystem is fed with two zero-rails and two one-rails, counting the number of computational basis states that result in detecting two photons in each subsystem is equivalent to counting the number of n-GHZ states.\\
For large $N$ it becomes impractical to expand the brackets of the generating function, and moreover, it would be nice to have a closed form expression for the coefficient of $\prod_{i=1}^Nu_i^2$, such that we can find a closed form expression for the number of n-GHZ states that yield two photons in each subsystem, a key step towards finding the success probability of the n-GHZ projection protocol. Luckily, for the connections $g_1$ and $g_2$, we find such closed form expressions, which we will discuss in the next subsection.

\subsection{Counting the number of candidate n-GHZ projections for $g1$}
We start by expanding the brackets in Eq. (\ref{eq: g1}) pairwise:
\begin{equation}
    g_1=\prod_{i=1}^N(u_i^2+2u_iu_{i+1}+u^2_{i+1}).
\end{equation}
We note that for each pair of qubits, we can either choose to send them both to the same subsystem ($u_i^2$ or $u^2_{i+1}$), or we can choose to send the qubits to two different subsystems in two ways ($2u_iu_{i+1}$). 

Let us consider the pair of qubits corresponding to $i=j$. If we choose to send both qubits to the same subsystem, let us say we pick $u^2_{j+1}$, then in the neighboring term ($i=j+1$), we are forced to pick $u^2_{(j+1)+1}$ since the other terms would result in three or more photons in subsystem $j+1$. This implies a chain reaction. If we choose the first term to be $u^2_{j+1}$, we are left with one choice for all other factors. The same argument holds if we choose the first term to be $u^2_{i}$. The contribution to the coefficient in front of $\prod_{i}u_i^2$ as a result of picking a square term is thus 2.

If we choose the first term to be $u_ju_{j+1}$, we are forced to pick 
$u_{j+1}u_{(j+1)+1}$ for $i=j+1$. Since otherwise we would either find three photons in subsystem $u_{j+1}$, or we would find merely one photon in subsystem 
$u_{j+1}$. Since we have always two ways to pick the 
$u_{j+1}u_{(j+1)+1}$, the contribution to the coefficient in front of $\prod_{i}u_i^2$ as a result of picking a cross term is thus $2^N$. The total coefficient, and therefore the number of n-GHZ states that we can potentially project on, is thus equal to $2+2^N$. Note that this is much larger than the 2 n-GHZ states we can potentially project onto using prior-art methods.

\subsection{Counting the number of candidate n-GHZ projections for $g2$}
We again start by expanding the brackets in Eq. (\ref{eq: g2}) pairwise:
\begin{equation}
    g_2=\prod_{i=1}^N(u^2_i+u_iu_{i+1}+u_iu_{i+2}+u_{i+1}u_{i+2}).
    \label{eq: g2 expanded}
\end{equation}
We note that this counting problem is a bit more tricky, because we have three possibilities for the qubits to split over two different subsystems. However, by careful analysis of considering all starting options we are able to find a closed form expression for the number of computational basis states (and thus also the number of n-GHZ) states that result in two photons detected in each subsystem. 

We realize that by considering three consecutive factors (e.g. $i=1$, $i=2$ and $i=3$) in the product of Eq. (\ref{eq: g2 expanded}), there is one subsystem that will not appear anymore in any of the other factors. That is, we must ensure that we have picked our terms in these three factors in such a way that the exponent of the dummy variable that represents this subsystem is equal to two. From here on, every next factor will contain one subsystem which will not appear in any of the other factors. Motivated by this observation, we will define "states" that represent the current status of the choices of terms made so far. These states can transform to other states by choosing a term in the next factor. However,  some transitions result into less or more than two photons in a subsystem that is not present in any other factors anymore, these transitions are therefore considered forbidden. In the remainder of this section, we will define more precisely what these states are, and what the transformation matrix looks like as a result of choosing a term in a factor from Eq. (\ref{eq: g2 expanded}). 

\begin{figure}
    \centering
    \includegraphics[width=0.9\linewidth]{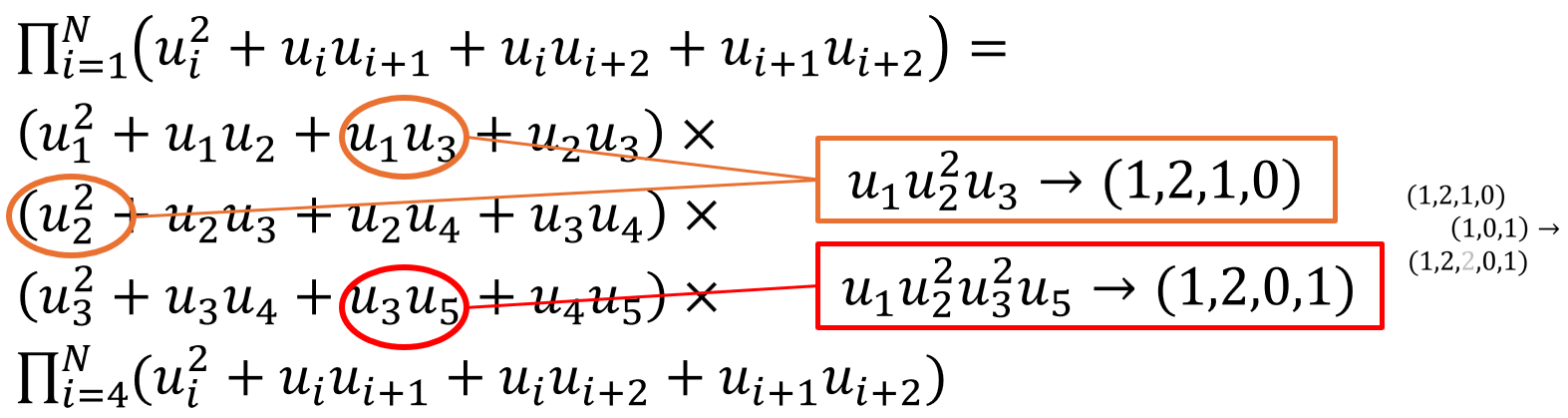}
    \caption{Example of a choice of terms and its corresponding state (orange). Example of a transition of a state (red).}
    \label{fig: defining states}
\end{figure}

We imagine choosing terms from the factors in Eq. (\ref{eq: g2 expanded}), starting by $i=1$ and continuing regularly, i.e. with $i=2$, $i=3$ etc.
We define states based on the value of the exponent of the first two and the last two subsystems, since these are subsystems that will appear in terms that are yet to come. All intermediate subsystems must be filled with exactly two particles, so we do not have to display these. The states are represented by a 4-tuple. Figure (\ref{fig: defining states}) illustrates an example of a choice of terms and the resulting state that is encoded in orange, and in red it shows how this state is updated after choosing a term in the next factor.

Each state will be a 4-tuple of which the elements are smaller or equal to two to ensure that there will not be more than two photons in one of the subsystems. The sum of the the elements of the 4-tuple will be equal to four for all states to ensure that none of the subsystems that will be present in any of the remaining factors will have less than two particles in it. All states are listed below.

\[
\begin{array}{rcl}
S_1 &=& (0,0,2,2)\\[4pt]
S_2 &=& (0,1,2,1)\\
S_3 &=& (0,1,1,2)\\[4pt]
S_4 &=& (1,0,2,1)\\
S_5 &=& (1,0,1,2)\\[4pt]
S_6 &=& (0,2,2,0)\\
S_7 &=& (0,2,1,1)\\
S_8 &=& (0,2,0,2)\\[4pt]
S_9 &=& (1,1,2,0)\\
S_{10} &=& (1,1,1,1)\\
S_{11} &=& (1,1,0,2)\\[4pt]
S_{12} &=& (2,0,2,0)\\
S_{13} &=& (2,0,1,1)\\
S_{14} &=& (2,0,0,2)\\[4pt]
S_{15} &=& (1,2,0,1)\\
S_{16} &=& (1,2,1,0)\\[4pt]
S_{17} &=& (2,1,0,1)\\
S_{18} &=& (2,1,1,0)\\[4pt]
S_{19} &=& (2,2,0,0)
\end{array}
\]

We can transform these states by choosing a term in the next factor. In each factor we have four choices: $u^2_i$, $u_iu_{i+1}$, $u_iu_{i+2}$ and $u_{i+1}u_{i+2}$, which we represent by the following 3-tuples respectively: $T_1=(2,0,0)$, $T_2=(1,1,0)$, $T_3=(1,0,1)$ and $T_4=(0,1,1)$. These 3-tuples correspond to the following qubits: $T_1=|00\rangle$, $T_2=|10\rangle$, $T_3=|01\rangle$ and $T_4=|11\rangle$. If a $S_i$ is updated according to transition $T$, the resulting state $S_f$ will be described by the following tuple: $(S_i(1),S_i(2),S_i(4)+T(2),T(3))$. Table (\ref{tab: transition table}) shows how each of the earlier mentioned 19 states are updated according to all four possible transitions. If an element has been left blank, the resulting state after the transition is not one of the 19 'valid' states since one of the subsystems that do not appear in any of the remaining factors is filled with either less than or more than 2 photons. 

\begin{table}[ht]
\centering
\small
\begin{tabular}{r|c c c c}
 & $T_1$ & $T_2$ & $T_3$ & $T_4$ \\\hline
$S_1$  &  &  &  &  \\
$S_2$  &  &  &  & $S_2$ \\
$S_3$  &  &  & $S_2$ &  \\
$S_4$  &  &  &  & $S_4$ \\
$S_5$  &  &  & $S_4$ &  \\
$S_6$  &  &  &  & $S_7$ \\
$S_7$  &  & $S_6$ & $S_7$ &  \\
$S_8$  & $S_6$ &  &  &  \\
$S_9$  &  &  &  & $S_{10}$ \\
$S_{10}$ &  & $S_9$ & $S_{10}$ &  \\
$S_{11}$ & $S_9$ &  &  &  \\
$S_{12}$ &  &  &  & $S_{13}$ \\
$S_{13}$ &  & $S_{12}$ & $S_{13}$ &  \\
$S_{14}$ & $S_{12}$ &  &  &  \\
$S_{15}$ & $S_{16}$ &  &  &  \\
$S_{16}$ &  & $S_{16}$ & $S_{15}$ &  \\
$S_{17}$ & $S_{18}$ &  &  &  \\
$S_{18}$ &  & $S_{18}$ & $S_{17}$ &  \\
$S_{19}$ & $S_{19}$ &  &  &  \\
\end{tabular}
\caption{Transformations of the 19 states $S_i$ by the four $T_j$.}
\label{tab: transition table}
\end{table}

If we look at table (\ref{tab: transition table}), we notice some interesting things. Firstly, we see that some states are never obtained after any transition (for example $S_8$). Secondly, we see that the states can only transform in certain cycles due to the fact that the first two elements of the tuple are unaffected by the transformation. 

We note that we can greatly simplify table (\ref{tab: transition table}) by considering the fact that not all states can be initialized, not all states can be transformed towards, and not all states can be closed.

By closing, we mean a sequence of two transitions, such that the periodic boundary conditions ensure that all subsystems contain two photons. For example, the state $S_7=(0,2,1,1)$ is closed with state $S_9=(1,1,2,0)$. 

\begin{table}[ht]
\centering
\small
\begin{tabular}{r|c c c c}
 & $T_1$ & $T_2$ & $T_3$ & $T_4$ \\\hline
$S_2$  &  &  &  & $S_2$ \\

$S_6$  &  &  &  & $S_7$ \\
$S_7$  &  & $S_6$ & $S_7$ &  \\

$S_9$  &  &  &  & $S_{10}$ \\
$S_{10}$ &  & $S_9$ & $S_{10}$ &  \\

$S_{15}$ & $S_{16}$ &  &  &  \\
$S_{16}$ &  & $S_{16}$ & $S_{15}$ &  \\
$S_{17}$ & $S_{18}$ &  &  &  \\
$S_{18}$ &  & $S_{18}$ & $S_{17}$ &  \\
$S_{19}$ & $S_{19}$ &  &  &  \\
\end{tabular}
\caption{Transformation of the states $S_i$ by the four $T_j$. We only consider transformation cycles that contain states that are innitializable and closable}
\label{tab: relevant transition table}
\end{table}

We recognize that there are two kinds of cycles: The first corresponds a cycle in which only one transition is allowed, and this transition moves the state back to itself (See $S_2$ or $S_{19}$). This means that if we are in one of these states, we will always be forced to choose a specific transition that transforms the state to itself. 

The second kind corresponds to a cycle in which three different transitions are allowed, featuring two states (see for example $S_6$ and $S_7$). One transition acts as a kind of identity transition, transforming one of the two states into itself. The other two transitions move between the two states. This means that whenever we are in one state, we can choose between two transitions, while if we are in the other state, we are forced to choose one specific transition.

We can recognize a combinatorial problem here. Given $K$ coins placed on a line, count the number of configurations, such that we never see two Heads in a row. The Heads and Tails here indicate the sequence of states that our system is in. For example: Heads is state $S_6$ and Tails is state $S_7$. Since there is no allowed transition that transforms state $S_6$ to state $S_6$, we will never see two states $S_6$ in a row. 

We can solve this problem by recognizing a linear recurrence relation. let $a_k$ be the solution to our problem for $k$ coins. Now, every sequence starts with either Heads or Tails. Say we have $N$ coins, if we start with Tails, the number of configurations for the remaining $N-1$ coins is equal to $a_{N-1}$. If we start with Heads, the second coin needs to be Tails and the number of configurations for the remaining $N-2$ coins is equal to $a_{N-2}$. Hence, we find the linear recurrence relation: $a_N=a_{N-1}+a_{N-2}$. For one and two coins, we can simply count the respective solutions to be $2$ (H or T) and $3$ (HT, TT or TH). Since the recurrence relation is equal to the one from the Fibonacci sequence and the initial conditions overlap with the Fibonacci sequence, the solution to our counting problem is equal to the Fibonacci numbers, albeit shifted ($a_N=F_{N+2}$).

The most straightforward way to continue is to count the number of ways we can initialize each state, count the number of ways to close each state and to count the number of different transitions from starting to closing states. However, there is a slightly more elegant way to continue. We realize that the choice for a starting position was completely arbitrary due to the periodic boundary conditions (We chose to start at $i=1$). Therefore, the initialization and closing of the states is also arbitrary, that is, the initialization and closing of states can just be considered as the bulk, if we would have chosen a different starting point. We thus realize that the transition cycles are the only thing that defines this problem. We do have to slightly adjust the problem, because due to the periodic boundary conditions, the coins should now be considered on a ring. We note there are four cycles we should take into account: 
\begin{enumerate}
    \item The cycle in which the allowed transitions are $T_1$.
    \item The cycle in which the allowed transitions are $T_4$.
    \item The cycle in which the allowed transitions are $T_1$, $T_2$ and $T_3$.
    \item The cycle in which the allowed transitions are $T_2$, $T_3$ and $T_4$.
\end{enumerate}

The number of computational basis states that result into two photons in each subsystem corresponding to the first two cycles is two (one for each) and corresponds to $|00\rangle^{\bigotimes N}$ and $|11\rangle^{\bigotimes N}$ respectively (remember, $N$ is the number of subsystems and half the number of qubits).

The number of computational basis states corresponding to the last two cycles is equal to the number of configurations of coins that are distributed on a ring, such that no two neighboring coins are both Heads. We can count this in the same way, let $b_k$ be the solution to this problem for $k$ coins. Consider the situation where we have $N$ coins, Without loss of generality, we pick one coin on the ring and say it can either be Heads or Tails. If it is Tails, the number of configurations for the remaining $N-1$ coins is equal to the number of configurations of $N-1$ coins on a \textit{line}, and hence equal to $a_{N-1}$. This is because the periodicity is "broken" by this Tails, even if on the remaining $N-1$ coins start and end with Heads, there is a Tails between them. If the first coin is Heads, then both of its neighbors must be Tails. The number of configurations for the remaining $N-3$ coins is equal to the number of configurations of $N-3$ coins on a line, and hence equal to $a_{N-3}$. So $b_N=a_{N-1}+a_{N-3}$, which is more famously known as the $N^{th}$ Lucas number ($L_N=F_{N+1}+F_{N-1}$).   

We conclude that the total number of computational basis states, and hence also the number of n-GHZ states, that distribute the photons such that there are two in each subsystem is equal to $2+2L_N$, or equivalently using Binet's forula: $2+2 (\phi^N+(1-\phi)^{-N})$, where $\phi\approx1.618$ is the golden ratio.

We note that this is exponentially increasing in the number of sybsystems (or qubits), but it grows slower than $2^N$. However as we will see in the next section, this connectivity will allow for more measurement patterns that herald a successful n-GHZ projection, ultimately resulting in a significantly larger success probability for a n-GHZ projection.

\subsection{Success probability of projecting onto a potential n-GHZ state}
We have seen that by considering more complex subsystems, the number of potential n-GHZ states that we can project on can be increased exponentially as the number of n-GHZ states for which we can successfully determine the sign of the X stabilizer generator is exponentially larger. However, we have carefully included the word 'potentially', since not every detection pattern that finds two photons in each subsystem will give us enough information about the Z generators to unambiguously determine which n-GHZ state has been projected on. We recall that an anti-bunched detection event in a subsystem will yield information about three pairwise Z generators that feature the four qubits that are linked to this subsystem. Each subsystem can either bunch or anti-bunch. Whenever a subsystem anti-bunches, we gain information about some Z generators of the state that has been projected on, when a subsystem bunches, we do not get such specific information.

The question thus becomes a counting problem in which the number of configurations are counted that yield a sufficient amount of information about the Z generators to determine exactly which n-GHZ state has been projected on. 

More specifically, we can view this as a graph connectivity problem. We consider a graph in which the vertices represent our qubits, and the edges represent pairwise Z stabilizers that feature the two qubits corresponding to the vertices it connects. Every time a subsystem ant-bunches, we have obtained three pairwise Z stabilizer generators, and hence we can add these edges to our graph. If and only if we have added a sufficient amount of edges such that our graph is connected, we can deduce all stabilizer generators. The problem thus becomes a counting problem where we count the number of different configurations of bunching and anti-bunching subsystems that result into a connected graph. 

We will consider the connections corresponding to Eqs. (\ref{eq: g1}) and (\ref{eq: g2}) separately. in the next subsections. After we have counted the number of configurations that result in a connected graph, this can be easily converted into a success probability for a n-GHZ projection. This is due to the fact that the probability for a subsystem to bunch is equal to the probability for a subsystem to anti-bunch, when an n-GHZ state is used as input. The probability of any configuration of bunching and anti-buncing subsystems to happen is $\textcolor{green}{\frac{2}{2^N}}$.

\subsection{Success probability for $g_1$}
The connectivity between qubits and subsystems described by $g_1$ implies that every time an anti-bunched detection pattern is observed after subsystem i ($i\in[N]=\{1,2,3, \dots\}$), information about the following three Z generators: $\{Z_{2i-1}Z_{2i}, Z_{2i}Z_{2i+1}, Z_{2i+1}Z_{2i+2} \}$ is obtained. (recall there are twice as many qubits as subsystems.). We note that information about some stabilizers appears in multiple subsystems ($Z_3Z_4$ is present for subsystem $i=1$ and $i=2$)\footnote{This might be considered as a feature, since it acts in some way as an additional round of measurements on some of the stabilizers, which could help detecting imperfections}, while others appear only in one subsystem (for example $Z_2Z_3$). This means that whenever subsystem $k$ anti-bunches, it disconnects the graph due to the lack of the edge corresponding to $Z_{2k}Z_{2k+1}$. However, we must not forget that we are dealing with periodic boundary conditions, hence one such disconnection is not a problem as long as all other subsystems bunch, the total graph will be connected. Since it does not matter which subsystem anti-bunches (or if they all anti-bunch), we find that the number of configurations of bunching and anti-bunching subsystems that result into a connected graph is $\textcolor{red}{N+1}$. \\
In total, there are \textcolor{cyan}{$2+2^N$} n-GHZ states that result into two photons in each subsystem. We consider \textcolor{cyan}{pairs} of n-GHZ states which have equal Z stabilizer generators. For each pair, there are $\textcolor{red}{N+1}$ out of $2^N$ configurations that yield a sufficient amount of information to deduce all Z stabilzers and the sign in front of the X stabilizer is simply deduced by counting the number of subsystems that anti-bunched. Hence, the success probability for a connection according to Eq. (\ref{eq: g1}) is equal to:
\begin{equation}
    P_{g_1}>\frac{\textcolor{cyan}{1+2^{N-1}}}{2^{2N}}\left((\textcolor{red}{N+1})\textcolor{green}{\frac{2}{2^N}}\right).
\end{equation}
However, there is one additional piece of information that we can still use that will increase this success probability. 

The subsystems that bunch may not tell us directly what the Z stabilizers are of the state we project onto, but they sometimes do so in an indirect way. If all subsystems contain two photons, we know that we are dealing with one of the $2+2^N$ n-GHZ states. If a subsystem anti-bunches such that the signs of all observed stabilizers is plus (by measuring either $|1010\drangle$ or $|0101\drangle$), we know for sure that we either projected onto $|0\rangle^{\bigotimes n}\pm |1\rangle^{\bigotimes n}$. This is because among the $2+2^N$ states that result into two photons in each subsystem, these are the only two states which can result into these anti-bunched outputs. This means that For the $|0\rangle^{\bigotimes n}\pm |1\rangle^{\bigotimes n}$ states, only observing one subsystem to anti-bunch is sufficient. There are \textcolor{blue}{$2^N-1$} such configurations. The other $2^N$ cannot be distinguished using less than $N-1$ anti-bunching subsystems in an indirect way. 
Hence, the total success probability for a connection according to Eq. (\ref{eq: g1}) is equal to:
\begin{equation}
    P_{g_1}=\frac{1}{2^{2N}}\left(\textcolor{cyan}{2^{N-1}}(\textcolor{red}{N+1})\textcolor{green}{\frac{2}{2^N}}+\textcolor{cyan}{1}\times(\textcolor{blue}{2^N-1})\textcolor{green}{\frac{2}{2^N}}\right).
\end{equation}

\subsection{Success probability for $g_2$}
Motivated by the fact that the connections between qubits and subsystems described by $g_1$ results in a lot of overlap in the potential information (For example $Z_3Z_4$ is present in two subsystems), we wanted to consider a different connection between qubits and subsystems. We considered the connections described by $g_2$. We have already seen that the number of n-GHZ states for which we can determine the sign of the X stabilizer is growing exponentially in the number of qubits ($\sqrt{\phi}^{n}$). Our aim is for the number of bunching and anti-bunching configurations that result in a sufficient amount of information to infer the Z stabilizers to grow exponentially too. This turns out to be the case for $g_2$.

\begin{figure}
    \centering
    \includegraphics[width=0.9\linewidth]{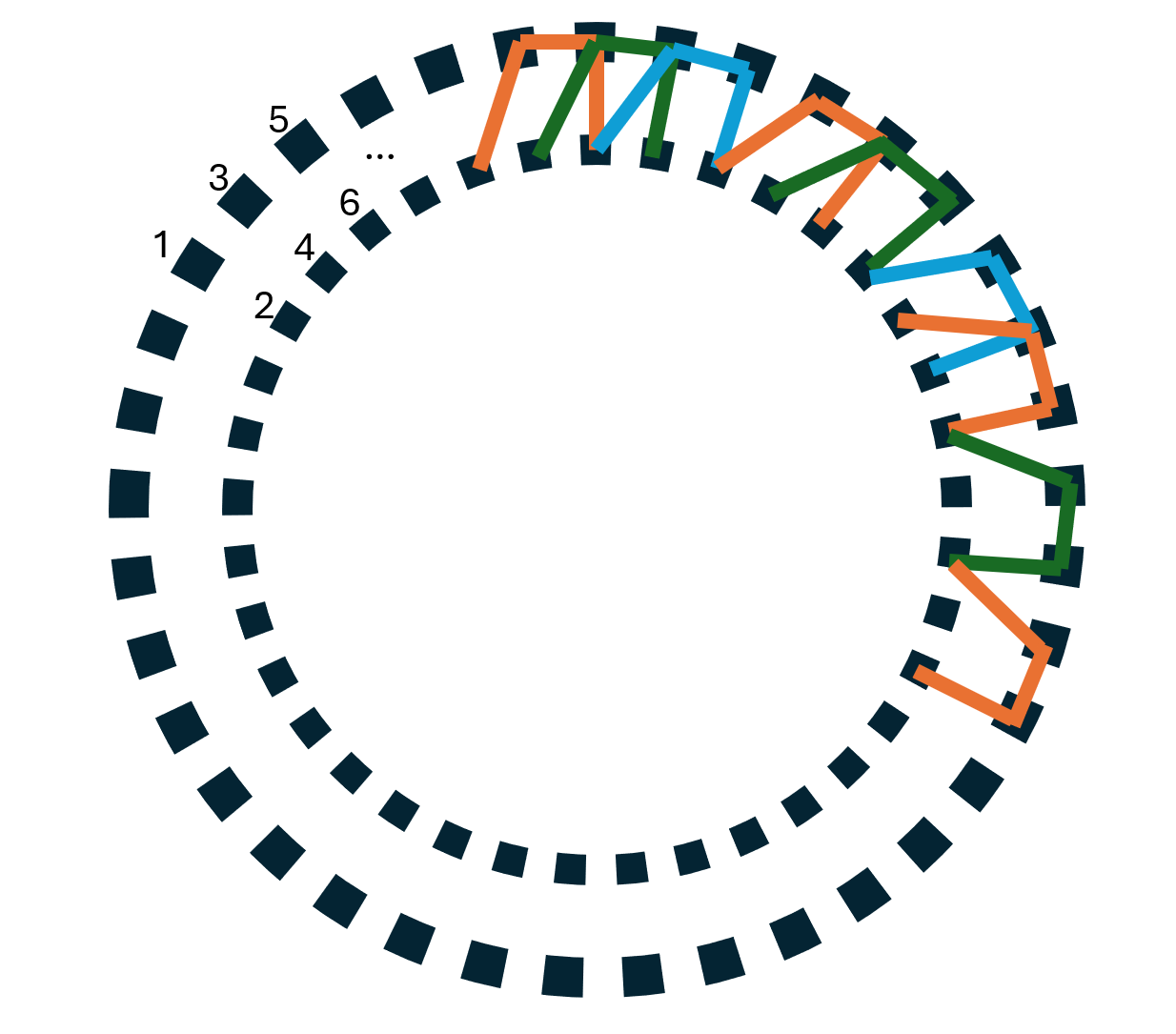}
    \caption{Graph in the shape of a double ring. Each vertex represents a qubit. Each edge represents a pairwise Z stabilizer. For the connections between qubits and subsystems according to $g_2$ from Eq. (\ref{eq: g2}), each anti-bunching subsystem results into a slanted-n shaped collection of edges. The collors are added for visual clarity and help identify which collection of edges is added as a result of an anti-bunching subssytem.}
    \label{fig: vertex connections g2}
\end{figure}

The connections between qubits and subsystems described by $g_2$ implies that every time an anti-bunched detection pattern is observed after subsystem i ($i\in[N]=\{1,2,3, \dots\}$), information about the following three Z generators: $\{Z_{2i}Z_{2i+1}, Z_{2i+1}Z_{2i+3}, Z_{2i+3}Z_{2i+4} \}$ is obtained. This more exotic gain of information allows for more subsystems to anti-bunch, while still resulting in a connected graph. In particular, a subsystem may bunch, as long as the two subsystems before and the two subsystems after it antibunch. Figure (\ref{fig: vertex connections g2}) illustrates why this is the case.

Figure \ref{fig: vertex connections g2} displays a graph in the shape of a pair of discretized rings. Each vertex represents a qubit, on the upper left part of the Fig. (\ref{fig: vertex connections g2}) is indicated a rule which tells us which vertex corresponds to which qubit. In orange, green and cyan are examples of added triplets of edges as a result of measuring an anti-bunching configuration in a certain subsystem. The colors are merely there to clarify which triplet of edges corresponds to which subsystem. 

In the upper right quadrant of the graph is displayed that a fully connected graph is obtained, even if a subsystem bunches, as long as the two subsystems before and after if anti-bunch. At the right of the Fig. (\ref{fig: vertex connections g2}) is displayed, that if only one system anti-bunches before another one bunches, there will be a vertex that is left unconnected. 

We can count the number of configurations that satisfy our constraint using similar techniques as discussed earlier. We would find a linear recurrence relation for the number of configurations on a \textit{line} where there are always at least two Tails between two Heads: $a_n=a_{n-3}+a_{n-1}$. We would then find that the number of configurations on a \textit{ring} can be expressed as $b_n=a_{n-2}+2*a_{n-5}$. We could continue to find a closed form expression for $b_n$ and we would find that it would scale as $\approx1.46557^N$. 

However, we do not want to do this calculation explicitly, because we will improve the success probability by considering indirect information, just like we did at the end of the previous subsection. I thought it would be still noteworthy that even when we do not consider the indirect information, this success probability scales better than the success probability for $g_1$. This is simply because for $g_2$ we find both an exponential increase in the number of n-GHZ states for which we can learn the X stabilizer, and for the number of bunching/anti-bunching configurations that allow us to determine the Z stabilizers. Moreover $1.46557\times\phi\approx2.371>2$. 

Just like in the case of $g_1$, the information that every subsystem contains two photons, in combination with information about some Z stabilizers will give us information about other Z stabilizers indirectly. However due to the more complex connections between qubits and subsystems, this is much more complex for $g_2$. We will continue to describe two ways in which additional information can be obtained. We know that there are more ways to obtain Z stabilizer information indirectly, but we did not manage to translate this to a closed form expression for the success probability, hence we will settle for a lower bound.

The first way to infer additional stabilizer information is again by considering the $|0\rangle^{\bigotimes n}\pm |1\rangle^{\bigotimes n}$ states. These are the only two n-GHZ states out of the $2+2L_n$ states (that result into two photons in each subsystem) that have four adjacent qubits such that all pairwise Z stabilisers are plus (a superposition of 0000 and 1111). This means that whenever we measure two adjacent subsystems such that the signs of all observed stabilizers is plus (by measuring twice either $|1010\drangle$ or $|0101\drangle$), we know for sure that we either projected onto $|0\rangle^{\bigotimes n}\pm |1\rangle^{\bigotimes n}$. Counting the number of configurations in which at least once are two adjacent subsystems that anti-bunch, is equivalent to counting the number of configuration in which never two adjacent subsystems anti-bunch in combination with counting the total number of configurations. Hence the number of bunching/anti-bunching configurations that allow us to successfully distinguish a projection onto $|0\rangle^{\bigotimes n}\pm |1\rangle^{\bigotimes n}$ equals \textcolor{blue}{$2^N-L_N$}.

We have seen in table \ref{tab: relevant transition table} that for computational basis states, the transformation updates are very limited, we either have two options, or the transformation is forced. This means that if we have a computational basis state that results into two photons in each subsystem, but we are missing some of the bits, we can often deduce what the missing bits must be due to the transformation rules. 

However, the situation we are interested in, is not determining a set of missing bits, but rather determining a set of missing pairwise Z stabilizers. Fortunately, for our situation, this is the same problem. If we look at table \ref{tab: relevant transition table} carefully, we note that seemingly different allowed transformation cycles can be described by describing transformation rules in terms of pairwise Z stabilizers. A transition cycle that contains $T_1=|00\rangle$, $T_2=|10\rangle$ and $T_3=|01\rangle$ (in which $T_1$ is a forced transition after $T_3$ and transitions $T_2$ and $T_3$ are possible after $T_2$ etc.) will have the result in allowed bitstrings which have the same pairwise Z stabilizers as a transition cycle that contains $T_2=|10\rangle$, $T_3=|01\rangle$ and $T_4=|11\rangle$. We tabulate the allowed pairwise Z stabilizers for a section in the bulk that contains six qubits.

\begin{table}[ht]
\centering
\small
\begin{tabular}{l|c}
\textbf{Computational basis states} & \textbf{Pairwise $Z_iZ_{i+1}$ stabilizers} \\[2pt]
\hline
$\ket{000000}, \ket{111111}$ & \texttt{+++++} \\
$\ket{101010}, \ket{010101}$ & \texttt{-----} \\
$\ket{001010}, \ket{110101}$ & \texttt{+----} \\
$\ket{010010}, \ket{101101}$ & \texttt{--+--} \\
$\ket{000100}, \ket{111011}$ & \texttt{++--+} \\
$\ket{100100}, \ket{011011}$ & \texttt{-+--+} \\
$\ket{010001}, \ket{101110}$ & \texttt{--++-} \\
$\ket{101001}, \ket{010110}$ & \texttt{---+-} \\
$\ket{001001}, \ket{110110}$ & \texttt{+--+-} \\
\label{tab: pairwize stabilizers for bulk}
\end{tabular}
\caption{Each pattern of pairwise $Z_iZ_{i+1}$ stabilizers corresponds to two computational basis states related by a global bit flip.}
\end{table}
After inspecting table \ref{tab: pairwize stabilizers for bulk} we note that only nine sequences are allowed, whereas without restrictions $2^5=32$ sequences would have been allowed. More specifically, we note that the sign of the second element is always equal to the sign of the fifth element. This means that we can deduce $Z_{i-1}Z_{i}$ as long as we know $Z_{i-4}Z_{i-3}$ and vice versa for all $i$ that are even. \footnote{Note that there are more correlations that can be deduced from table \ref{tab: pairwize stabilizers for bulk}, but these are specific for certain n-GHZ states and are therefore more difficult to include. One of such correlations is that if $Z_{i-1}Z_i=+1$ for some $i$ that is even, and if any other pairwise correlation is equal to $-1$, then it follows that $Z_{i+1}Z_{i+2}=-1$ and $Z_{i-3}Z_{i-2}=-1$. For relatively small system sizes, it is feasible to numerically compute the success probability exactly, taking into account all correlations that are present.}

\begin{figure}
    \centering
    \includegraphics[width=1\linewidth]{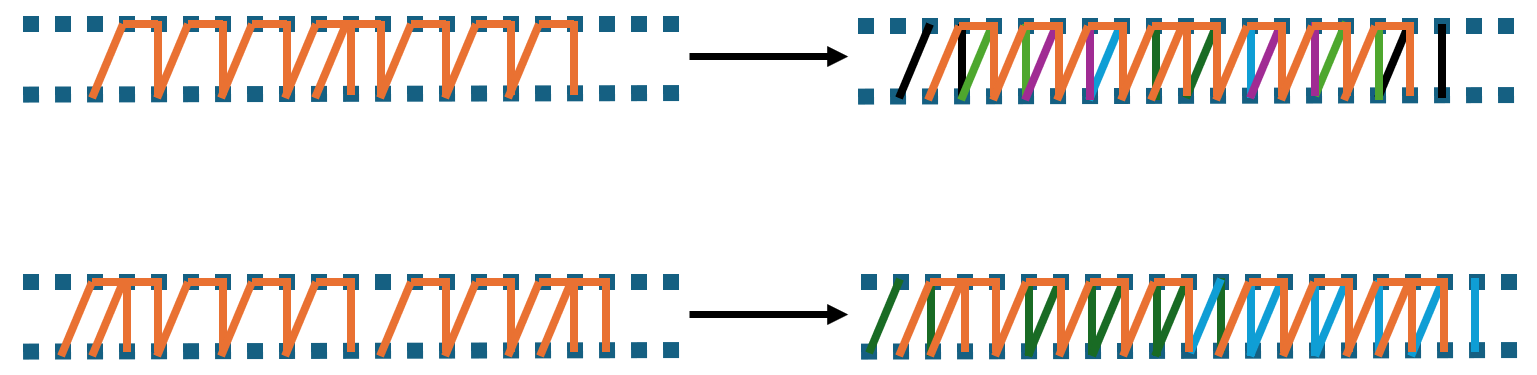}
    \caption{We display the bulk section of a double ring graph as presented in Fig. (\ref{fig: vertex connections g2}). The figures on the left indicate the stabilizer information (in orange) that is obtained as a direct result of anti-bunching susbsystems. The figures on the right indicate the stabilizer information that can indirectly be deduced by multiplying certain known stabilizers and applying the restriction imposed by the limited number of allowed transitions. The top figures shows a sequence of subsystems that bunch and anti-bunch in an alternating fashion, while in the center there are two neighboring subsystems that anti-bunch. In the top right figure we show how the neighbouring anti-bunching pair imposes a chian reaction of additional stabilizer information. The colors indicate which stabilizers are deduced before others, following this temporal ordering: \textcolor{OliveGreen}{dark green}, \textcolor{Cyan}{cyan}, \textcolor{Plum}{purple}, \textcolor{LimeGreen}{light green} and \textbf{black}. The bottom figures show a sequence of of subsystems that bunch and and anti-bunch in an alternating fashing, with a gap in the center, where two neighbouring subsystems bunch. Additionally at both end points, two neighbouring systems anti-bunch. In the figure on the bottom right is indicated which additional stabilizer information can be deduces as a result of each ati-bunching pair. We note that we can 'fill' the gap created by two neighbouring bunching subsystems whenever it is enclosed by two instances of a pair of neighboring anti-bunching pairs.}
    \label{fig:placeholder}
\end{figure}

If we look at Fig. \ref{fig: vertex connections g2} it is not immediately clear that this is an advantage. More specifically, each anti-bunching event gives us information about three Z stabilizers, a diagonal, a horizontal and a vertical one. The diagonal and the vertical exactly correspond to $Z_{i-1}Z_{i}$ and $Z_{i-4}Z_{i-3}$ for all $i$. However, there is a way to indirectly obtain information about either $Z_{i-1}Z_{i}$ or $Z_{i-4}Z_{i-3}$. If we observe two neighboring subsystems to anti-bunch (say subsystems 3 and 4) we observe the following stabilizer pairs: $Z_2Z_3$, $Z_3Z_5$, $Z_5Z_6$, $Z_4Z_5$, $Z_5Z_7$ and , $Z_7Z_8$. From these stabilizers we can deduce the signs of $Z_3Z_4$ and $Z_6Z_7$, simply by multiplying the right two pairs. Now we can use the rule we just described to additionally deduce $Z_nZ_1$ and $Z_9Z_{10}$. Hence, two anti-bunching subsystems will yield more information due to the restrictive transformations. Note that in our example $Z_nZ_1$ or $Z_9Z_{10}$ could in turn be used to deduce more information. In general, we find that if two neighboring subsystems anti-bunch, we are able to 'fill' the disconnections that could be the result of subsystems that bunch. In fact, we are even able to 'fill' the disconnections that are a result of two neighboring systems that bunch, as long as two neighboring systems anti-bunch before another neighboring pair of subsystems bunches. See the Fig. (\ref{fig:placeholder}) for an illustration. 

It is quite difficult to count the number of configurations in which at most two neighboring subsystems bunch, such that between two of these neighboring bunching pairs, there is at least once a sequence of at least two neighboring anti-bunching subsystems. The reason for this to be difficult is that there is a global constraint. We have not been able to solve this, hence we continue counting the number of configurations in which no two neighboring subsystems bunch and at least once two neighboring subsystems anti-bunch. This counting problem is one that we have practically already solved. It essentially equals the number of configurations of coins on a ring where there are never two neighboring Heads, which we have seen to be equal to the Lucas numbers. However, there is a small caveat when the total number of coins is even. Then two configurations that never have two adjacent Heads, but also not have at least one instance of two adjacent Tails is present. These configurations are alternating Heads and Tails (..HTHTH..), and the same configuration but shifted by one. Hence the total number of configurations in which no two adjacent subsystems bunch and there is at least one instance of two adjacent subsystems to anti-bunch is equal to \textcolor{red}{$\tilde{L}_N$}. Here $N$ is the number of subsystems and \[
\tilde{L}_N = 
\begin{cases}
L_N, & \text{if $N$ is odd},\\
L_N -2, & \text{if $N$ is even.}
\end{cases}
\]
We recall that there are in total \textcolor{cyan}{$2+2L_N$} n-GHZ states (and thus \textcolor{cyan}{$1+L_N$} \textcolor{cyan}{pairs}) that result in two photons in each subsystem. A projection on \textcolor{cyan}{one} of these \textcolor{cyan}{$1+L_N$} pairs only requires an anti-bunching event in two adjacent subsystems, for which there are \textcolor{blue}{$2^N-L_N$} different configurations. While a projection on one of the other \textcolor{cyan}{$L_N$} pairs at most requires one pair of adjacent anti-bunching subsystems while there is no adjacent pair of bunching subsystems, for which there are \textcolor{red}{$\tilde{L}_N$} different configurations. Each configuration occurs with a probability of \textcolor{green}{$\frac{2}{2^N}$}. Hence, a lower bound on the success probability of an n-GHZ projection for a connection according to Eq. (\ref{eq: g2}) is equal to:

\begin{equation}
    P_{g_2}>\frac{1}{2^{2N}}\left(\textcolor{cyan}{L_N}(\textcolor{red}{\tilde{L}_N})\textcolor{green}{\frac{2}{2^N}}+\textcolor{cyan}{1}\times(\textcolor{blue}{2^N-L_N})\textcolor{green}{\frac{2}{2^N}}\right).
    \label{eq: Psucc g2}
\end{equation}
Especially interesting about Eq. (\ref{eq: Psucc g2}) is that for a large number of qubits, both the number of n-GHZ states for which we can determine the sign in front of the X stabilizer as well as the number of bunching/anti bunching configurations grow exponentially with $N$. As a result, the fraction of this new success probability compared to the success probability of the prior art grows exponentially.

\end{document}